\documentclass[usenatbib,twocolumn]{mn2e}
\usepackage[]{graphicx}
\usepackage[]{amssymb}

\def\erg{{\rm\thinspace erg}}
\def\s{{\rm\thinspace s}}

\def\ergps{{\rm \erg\s^{-1}}}

\def\zfac{{F_{\rm z}}}

\title[Are  GRBs the same at high-$z$ and low-$z$?]{Are gamma-ray bursts the same at high redshift and low redshift?}
\author[O.M. Littlejohns et al.]
{O.M. Littlejohns$^{1,2}$,\thanks{E-mail: olittlej@asu.edu (OML)}
N.R. Tanvir$^{1}$,
R. Willingale$^{1}$,
P.A. Evans$^{1}$,
P.T. O'Brien$^{1}$,
\and
A.J. Levan$^{3}$
\\
$^{1}$ Department of Physics and Astronomy, University of Leicester,
 Leicester, LE1~7RH, UK\\
$^{2}$ School of Earth and Space Exploration, Arizona State University, Tempe, AZ, 85287, USA\\
$^{3}$ Department of Physics, University of Warwick, Coventry,
 CV4 7AL, UK\\
}
\begin{document}
\date{2013 Sep 9}
%\date{Accepted 2005 June 15. Received 1988 December 14; in original form 1988 October 11}

\pagerange{\pageref{firstpage}--\pageref{lastpage}} \pubyear{2012}

\maketitle

\label{firstpage}
\begin{abstract}
The  majority  of {\em  Swift}  gamma-ray  bursts  (GRBs) observed  at
$z\gtrsim6$ have prompt  durations of $T_{90}\lesssim30$\,s, which, at
first sight, is surprising given that cosmological time-dilation means
this  corresponds to  $\lesssim5$\,s in  their rest  frames.   We have
tested whether the high-redshift  GRBs are consistent with being drawn
from  the  same  population  as  those  observed  at  low-redshift  by
comparing  them to  an artificially  red-shifted sample  of  114 $z<4$
bursts.  This  is accomplished using two methods  to produce realistic
high-$z$   simulations  of   light  curves   based  on   the  observed
characteristics of  the low-$z$ sample.  In  Method 1 we  use the {\em
  Swift}/BAT data directly, taking  the photons detected in the harder
bands to predict what would be  seen in the softest energy band if the
burst were  seen at higher-$z$.  In  Method 2 we fit  the light curves
with a model, and use  that to extrapolate the expected behaviour over
the whole BAT  energy range at any redshift.  Based  on the results of
Method 2, a K-S test  of their durations finds a $\sim$1\% probability
that the  high-z GRB sample is  drawn from the same  population as the
bright  low-z sample. Although  apparently marginally  significant, we
must bear in mind that  this test was partially \textit{a posteriori},
since the rest-frame short durations of several high-\textit{z} bursts
motivated the study in the first instance.
\end{abstract}


\section{Introduction}
\label{sec:intro}

Gamma--ray  bursts (GRBs)  are identified  as short-lived  (seconds to
minutes   in    most   cases)   transient    flashes   of   gamma-rays
\citep{1973ApJ...182L..85K}.  Their light  curves show great diversity
in  behaviour, ranging  from the  very  smooth to  the highly  erratic
\citep{2012ApJS..199...18P,2011ApJS..195....2S,2008ApJS..175..179S,1999ApJS..122..465P}.
Much effort has  been expended over the decades  in trying to classify
and               understand               this              diversity
\citep{2013ApJ...764..179B,2011ApJ...734...96K,2007ApJ...655L..25Z,2006ApJ...643..266N,2006Natur.444.1044G,2001ApJ...552...57R,2000ApJ...534..248N,1993ApJ...413L.101K},
although arguably the most useful observables remain the comparatively
gross  properties  of  duration  and average  spectral  hardness.   In
particular, these  properties serve to separate out  the two classical
sub-classes  of GRB,  namely the  long-duration/soft-spectrum  and the
short-duration/hard-spectrum
\citep{2013ApJ...764..179B,1993ApJ...413L.101K}.\par

An important question  is whether the populations of  GRBs change with
redshift, which in principle might  be reflected in the typical prompt
behaviour.    In   fact,   it    does   seem   that   the   short-GRBs
\citep{2007PhR...442..166N}  are  on average  fainter  and visible  at
lower redshifts  than the long-GRBs.  Beyond this,  the only tentative
evidence  for an  evolution in  the  population of  long-GRBs is  that
several  of the highest  redshift GRBs  found to-date  have apparently
rather  short durations,  $T_{90}\lesssim5$\,s, in  their  rest frames
\citep{2012MNRAS.421.1874G,2011ApJ...736....7C,2009Natur.461.1254T,2009ApJ...693.1610G}.
Several  works  have argued  that  these  are  most likely  {\em  not}
misclassified                                              short-bursts
\citep[e.g.,][]{2012arXiv1211.1117L,2009ApJ...703.1696Z,2010ApJ...708..117B},
so  it is natural  to ask  whether their  unusual properties  could be
indicating some  change in the typical  long-GRB progenitors, possibly
due to their having very low metallicity.  The difficulty in assessing
the significance of this finding is firstly that samples remain rather
small, and  secondly that measured  duration is actually  dependent on
detector  sensitivity and  band-pass,  in addition  to the  underlying
behaviour  of  the  given   GRB  and  indeed  the  chosen  operational
definition of ``duration".  Hence even when using the same instrument,
inferring rest-frame duration by simply dividing the observed duration
by the  cosmological time-dilation factor may  well produce misleading
results.\par

In a previous study \citet{2013ApJ...765..116K} simulated GRBs as they
would be observed by the Burst and Transient Source Experiment (BATSE;
\citealt{1992Natur.355..143M}) instrument,  to find whether signatures
of time dilation might be detected in properties such as $T_{90}$ (see
\S \ref{duration}).  In that work  the authors used  prescriptions for
the shape  and time evolution  of GRB spectra to  produce single-pulse
prompt  high-energy  light   curves.  These  simulations  showed  that
cosmological  time dilation  is often  not reflected  in  the measured
duration of  a burst.  In some  instances the duration  of a synthetic
burst  could  be seen  to  decrease as  a  function  of the  simulated
redshift, particularly when the signal  to noise ratio became poor.  A
limitation  of the  work presented  in  \citet{2013ApJ...765..116K} is
that  the   simulated  light  curves  all  contained   only  a  single
morphological feature:  one Fast Rise Exponential  Decay (FRED) pulse.
The authors suggested that time dilation might be more apparent in GRB
temporal profiles that contained narrow pulses separated by periods of
quiescence.  In such a situation,  it would be the increasing duration
between  peaks   that  would  provide  the  signature   of  this  time
dilation.\par

In this  paper we seek  to make a  more robust comparison  of observed
low- and  high-redshift GRB populations,  by taking a large  sample of
the former and artificially ``redshifting"  them to see how they would
appear if they  had occurred at high-$z$.  This  procedure is amenable
to simulation  since signal-to-noise  tends to reduce  with increasing
redshift,   largely   masking    uncertainties   introduced   in   the
band-shifting.   We  develop  two  methods of  simulation,  which  are
presented in \S~\ref{sec:modelling}, along  with our recipe to emulate
the      \textit{Swift}      Burst      Alert     Telescope      (BAT;
\citealt{2005SSRv..120..143B})  ``rate triggering''  (only) algorithm,
and  descriptions of the  various duration  measures we  consider.  In
\S~\ref{sec:disc} we study in detail the evolution of 16 bright bursts
with  simulated redshift.  Finally  in \S~\ref{sec:compks}  we analyse
the simulated prompt light curves  of 114 ($z<4$) GRBs and compare the
detectable   fraction    to   the   currently    known   high-redshift
\textit{Swift}  bursts,  assessing whether  they  are consistent  with
being drawn from the same parent population.\par

\section {Measures of duration}
\label{duration}

\subsection{$T_{90}$}
\label{sec:t90expl}

\begin{figure}
 \hspace{-2mm}\includegraphics[width=9.2cm,angle=0,clip]{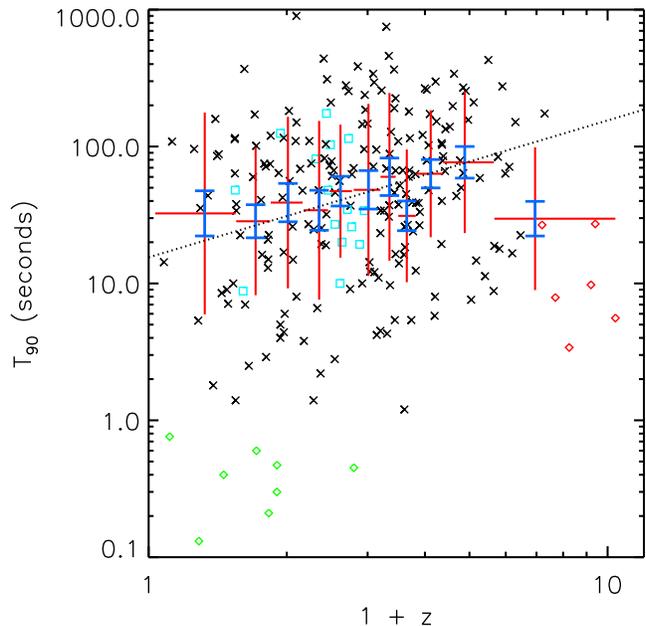}
  \caption{The   observed  distribution  of   $T_{90}$  for   the  203
    \textit{Swift} bursts with a  redshift estimate prior to 2012 July
    15$^{{\rm  th}}$  (note  this  includes  a  mixture  of  redshifts
    determined directly  from afterglows  and those inferred  via host
    galaxy   spectroscopy).   Also   included  are   GRB~120521C   and
    GRB~120923A which both are  candidates for high-redshift.  The red
    crosses show geometric averages of  bursts taken with 20 bursts in
    each bin (except  in the final bin which has  16 bursts).  The red
    error  bars shown  in the  vertical direction  show the  root mean
    square (RMS)  scatter calculated logarithmically.   Also shown are
    the standard errors  on the mean for each bin  in blue. The dotted
    black line shows the expected evolution due to simple cosmological
    time dilation,  namely $T_{90}\propto1+z$. Short  GRBs are denoted
    by the green diamonds.  The high-redshift subset of six GRBs shown
    in  Figure  \ref{fig:lchiz} are  indicated  by  red diamonds.  The
    low-redshift, bright  subset introduced in  \S \ref{sec:selbright}
    are  denoted   by  blue  squares.  Redshift   were  obtained  from
    http://swift.gsfc.nasa.gov/docs/swift/archive/grb\_table/.}
  \label{t90_vs_z}
\end{figure}

The durations of the high--energy prompt emission of GRBs are commonly
parameterised by $T_{90}$  which is defined as the  time between which
5\% and  95\% of the total  fluence received in  the observer-frame is
measured (similarly  $T_{50}$, for example, is the  time between which
25\%  and  75\%  of  the   total  fluence  is  received).   In  Figure
\ref{t90_vs_z} we  plot the  $T_{90}\left( 15 -  350~{\rm keV}\right)$
distribution as  a function of redshift  for the 203  GRBs detected by
\textit{Swift} \citep{2004ApJ...611.1005G} since launch that were seen
prior to  the 2012 July 15$^{{\rm  th}}$ and have  a measured redshift
(either  from  emission  lines,  absorption  lines  or  a  photometric
redshift).   We  use  the  values  for  $T_{90}$  available  from  the
\textit{Swift}  ground analysis,  which  are obtained  by running  the
Bayesian  blocks  algorithm  {\sc  battblocks}  as part  of  the  {\sc
  batgrbproducts} script.  {\sc battblocks}  is run on light curves of
several  different bin  sizes (4  ms, 16  ms, 1  s and  16  s), before
applying a  set of  criteria to determine  the best  $T_{90}$ duration
estimate \citep{2011ApJS..195....2S}.\par

The calculated means do not include short GRBs\footnote{Classically, a
  short  GRB has  usually been  taken to  be one  with $T_{90}$  $<$ 2
  seconds  \citep{1993ApJ...413L.101K}.  However, this  demarcation is
  valid (in the  sense of the long and  short populations contributing
  roughly 50:50  at this point) for  the BATSE whereas  in general one
  expects  it  to depend  on  the  instrument  and energy  band  used.
  Through   a  recent  analysis   of  the   \textit{Swift}  population
  \citet{2013ApJ...764..179B} conclude that for bursts detected by BAT
  the division should be at 0.8 seconds, and it is this value we adopt
  in Figure \ref{t90_vs_z}.}, as these tend to be observed in the more
local  Universe.  As  such  they  reduce the  averages  of the  lowest
redshift  bins   in  a  manner   that  could  be  mistaken   for  time
dilation.\par

Whilst the geometric RMS scatter in Figure \ref{t90_vs_z} is large, it
is interesting that  the mean values can be  seen to increase slightly
over  almost the  entire redshift  range of  the  observed population,
broadly in line with the expected effect of cosmological time dilation
i.e.,  $T_{90}\propto1+z$.   The standard-errors  on  these means  are
shown as shorter error bars, suggesting that the trend is no more than
moderately significant, and, indeed,  that a no-evolution model cannot
be  ruled  out.  To  test  this  we fitted  two  models:  one with  no
evolution,  finding  $T_{90}=45.8\pm5.4$  seconds, and  another  where
$T_{90}\propto1+z$  ($T_{90}  =  11.7\pm1.6  \left(  1  +  z  \right)$
seconds). For  the no-evolution model we  obtained a $\chi^{2}=10.39$,
for  10   degrees  of  freedom   ($\chi^{2}_{\nu}=1.04$),  whilst  the
$T_{90}\propto1+z$  model  had  a  $\chi^{2}=5.93$ for  9  degrees  of
freedom (reduced $\chi^{2}$=0.66).  Of the two, the $T_{90}\propto1+z$
has   the   lower   $\chi^{2}_{\nu}$   fit   statistic,   by   $\Delta
\chi^{2}_{\nu}=0.38$.  Only for the  final bin, containing the highest
redshift bursts  with $4.65$  $\leqslant$ $z$ $\leqslant$  $9.4$, does
the geometric mean $T_{90}$ fall significantly below either trend.\par

\begin{figure}
  \begin{center}
    \includegraphics[angle=0,width=9.2cm,clip]{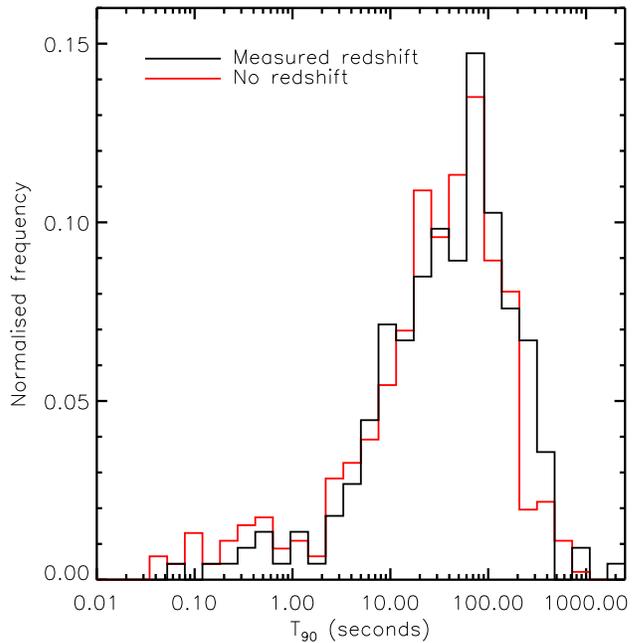}
  \end{center}
  \caption{$T_{90}$ distributions  for \textit{Swift} bursts  with and
    without  redshifts. The black  line corresponds  to bursts  with a
    measured redshift, whilst the red distribution is for those bursts
    which  don't. The  total  population extends  from  the launch  of
    \textit{Swift},  including  224 bursts  with  a  redshift and  459
    without one.}
  \label{fig:obst90hist}
\end{figure}

Shown  in  Figure   \ref{fig:obst90hist}  are  the  distributions  all
\textit{Swift} burst $T_{90}$ values when divided according to whether
a  redshift  was  measured  for   each.   As  can  be  seen,  the  two
distributions agree with  one another to a reasonable  extent. To test
this  we  performed  a   Kolmogorov-Smirnov  (K-S)  test  on  the  two
samples. This  statistical test compared  the cumulative distributions
of two  data sets to find whether  the two arise from  a common parent
population. From  the K-S test,  we derived a probability  of $p=0.1$,
which is  large enough  that the two  samples cannot  be distinguished
with  any  statistical  significance.    This  implies  that  the  two
populations  shown  in Figure  \ref{fig:obst90hist}  are not  markedly
different and that no duration  bias is introduced by only considering
bursts with available redshifts.\par

\subsection{An alternative measure of duration}
\label{sec:deftc}

\begin{figure}
  \begin{center}
 \hspace{-2mm}\includegraphics[width=9.2cm,angle=0,clip]{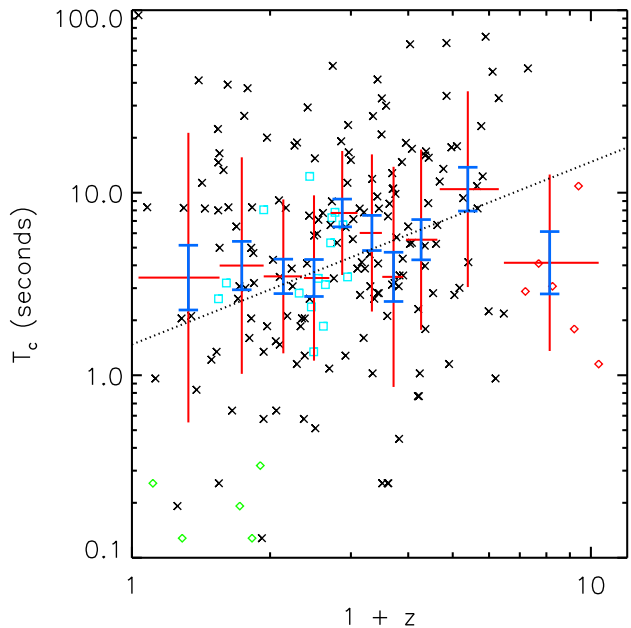}
  \end{center}
  \caption{The   observed   distribution   of   $T_{\rm   c}$   versus
    redshift. The high-redshift burst sample are shown by red diamonds,
    whilst the  low redshift sample  are shown by blue  squares. Short
    bursts  ($T_{90} < 0.8$  seconds) are  denoted by  green diamonds.
    The  dashed  line shows  the  $T\propto1+z$  expected from  simple
    cosmological    time dilation.     As    with   $T_{90}$    (Figure
    \ref{t90_vs_z}) the binned averages show a weak trend of increasing
    duration with redshift up to $z\sim5$.}
  \label{tc_vs_z}
\end{figure}

In  an attempt  to diminish  the effect  of background  noise  we also
adopted  an alternative measure  of prompt  duration.  From  a dynamic
binning  routine  that  bins the  64  ms  light  curves to  a  minimum
significance threshold,  we found the  brightest bin in  the re-binned
light curve.  We  used this value to define  a brightness threshold of
half of this maximum flux ($F_{\rm thresh}$ $=$ 0.5$F_{\rm max}$).  We
summed the duration of all  bins which were above this threshold. This
was termed  the ``core time" $T_{\rm  c}$, and measures  the period of
time   over  which   the  source   can  be   considered  to   be  most
active. Measures  such as $T_{90}$, which are  defined over contiguous
bins,  may  include quiescent  periods  between  regions of  activity,
whereas $T_{\rm c}$ will remain  insensitive to such times.  A plot of
$T_{\rm  c}$ versus  redshift is  shown in  Figure  \ref{tc_vs_z}, and
again demonstrates a weak  trend of increasing duration with redshift,
up to $z\sim5$.   A similar measure has previously  been used in works
such as \citet{2001ApJ...552...57R}:  specifically they considered the
duration of the  brightest bins in the light  curve that accounted for
50\% of the observed fluence.\par

As   for    $T_{90}$,   we   fitted   both    a   no-evolution   model
($T_{c}=5.23\pm0.38$ seconds) and one  which was proportional to $1+z$
($T_{c}  = 1.30\pm0.11(1+z)$ seconds),  as expected  from cosmological
time dilation.  For the no-evolution model, we found $\chi^{2}=30.26$,
with 9  degrees of freedom ($\chi^{2}_{\nu}=3.36$),  whilst the $T_{c}
\propto \left(1 + z \right)$  model had $\chi^{2}=26.25$ for 8 degrees
of freedom ($\chi^{2}_{\nu}=3.28$).  In terms of $\chi^{2}_{\nu}$, the
$T_{c} \propto \left(1  + z \right)$ proved a  better fit with $\Delta
\chi^{2}_{\nu}=0.08$, although this improvement is marginal due to the
nature of the  data. The $T_{c} \propto \left(1 +  z \right)$ model is
the dotted line plotted in Figure \ref{tc_vs_z}.\par

\section{Data}
\label{sec:obs}

As a basis for our  simulations of high-redshift bursts, we considered
GRBs observed by  \textit{Swift} prior to 2011 May  3$^{{\rm rd}}$ for
which redshifts have been reported.

\subsection{Selection of low-redshift bursts}
\label{sec:selbright}

This work  aims specifically to predict the morphology of  the prompt
high-energy  light curves,  and key  parameters such  as  $T_{90}$, of
high-redshift GRBs  based on the properties of  those at low-redshift.
To  maximise the  baseline in  redshift,  we considered  only the  175
\textit{Swift} bursts  prior to 2011 May 3$^{{\rm  rd}}$ with observed
redshift, $z_{\rm  orig}$ $<$ 4.   Of these, we retained  those fitted
with  the \citet{2010MNRAS.403.1296W}  pulse model  (essentially these
were all {\em Swift} GRBs observed before 2011 May 3$^{{\rm rd}}$, for
which the satellite was able to execute a prompt slew and hence gather
early X-ray data; see also \S \ref{sec:xrtobs}). This yielded 114 GRBs
which   were   simulated  at   high-redshift,   as   detailed  in   \S
\ref{sec:compks}.\par

To  also effect a  detailed case  study of  the evolution  of measured
durations with  increased simulated redshift, we  identified a smaller
sample of bursts,  bright enough to have a  realistic possibility that
they would have been detected (see also \S~\ref{sec:ratealg}) had they
occurred  at $z\sim6$.   We identified  these as  having at  least one
pulse with a  peak bolometric luminosity $L_{\rm pk}$  $>$ $1.0 \times
10^{52}\ergps$. This value was chosen  to ensure that we simulated all
bursts  that  produced  detectable  structure  using  both  simulation
methods at $z_{{\rm  sim}} \sim 6$, but to  also avoid the simulations
of  a  larger  sample  of   bursts  that  would  not  be  detected  at
high-redshifts.\par

This  peak  luminosity  made  use  of  the  model  fits  described  in
\S~\ref{sec:pulsim},    and     was    calculated    using    Equation
\ref{eq:lpkcalc}, where $D_{L,orig}$ is the luminosity distance to the
burst  as  observed, $S_{pk}$  is  the  brightest pulse  normalisation
across  the BAT  band  as defined  in \citet{2010MNRAS.403.1296W}  and
$K_{\rm orig}$  is the $K$-correction  required to convert this  15 --
350 keV fluence normalisation to a bolometric fluence.\par

\begin{equation}
  L_{\rm pk} = 4 \pi D_{\rm L,orig}^{2}  K_{\rm orig} S_{\rm pk} .
  \label{eq:lpkcalc}
\end{equation}

This  selection produced  16 bursts,  some with  multiple  pulses that
satisfied the  luminosity threshold. These GRBs and  the properties of
their brightest pulses are  detailed in Table \ref{tab_bright}.  It is
important  to  emphasise that  since  we  are  dealing with  just  the
brightest $z_{{\rm  orig}}<2$ GRBs it  is unlikely that  any selection
effects, such  as the requirement  for a redshift  determination, will
introduce any bias in the distribution of durations.\par

\begin{table*}
  \centering
  \caption{  The  bright  GRB   subset  identified  for  our  detailed
    morphological case  study.  Values of $T_{90}$  are those obtained
    from  the standard  BAT analysis.  $N_{p}$ denotes  the  number of
    modelled  pulses used  for  fitting the  light curves,  $b_{1}$,
    $S_{pk}$  and $L_{pk}$ all  correspond to  the relevant values of the brightest  pulse in
    each      burst.       (Redshifts      obtained      from:      1.
    \citet{2005GCN..3483....1F},  2.   \citet{2007A&A...468...83V}, 3.
    \citet{2009ApJS..185..526F},   4. \citet{2007ApJ...663.1125P},   5.
    \citet{2008Natur.455..183R},  6.   \citet{2008GCN..7517....1W}, 7.
    \citet{2012A&A...546A...8K},  8.   \citet{2010MNRAS.405.2372G}, 9.
    \citet{2009GCN..9243....1C}, 10.  \citet{2010A&A...516A..71M}, 11.
    \citet{2009GCN..10053...1X},   12.   \citet{2010GCN..11089....1O},
    13. \citet{2012MNRAS.421.1874G}, 14.  \citet{2011ApJ...743..154C},
    15.                 \citet{2011GCN..11978...1D}                and
    16. \citet{2011GCN..11997...1D})}
  \label{tab_bright}
  \begin{tabular}{@{}ccccccc}
    \hline
    \hline
    GRB & Redshift & $T_{90}(z_{\rm orig})$ & $N_{\rm p}$ & $b_{1}$ & $S_{\rm pk}$ &
    $L_{pk}$ \\
    & & (s) & & & (keV\,cm$^{{\rm -2}}\,{\rm s}^{-1}$) & ($\ergps$) \\
    \hline
    GRB~050525A & 0.606$^{{\rm 1}}$ & 8.86 $\pm0.07$ & 7 & 0.06 &
    4441.78 & 1.08 $\times$ 10$^{{\rm 52}}$ \\
    GRB~060418 & 1.489$^{{\rm 2}}$ & 109.17 $\pm46.73$ & 9 & -0.38 &
    576.42 & 1.32 $\times$ 10$^{{\rm 52}}$ \\
    GRB~060908 & 1.884$^{{\rm 3}}$ & 18.78 $\pm1.30$ & 4 & 0.11 &
    256.88 & 1.05 $\times$ 10$^{{\rm 52}}$ \\
    GRB~061121 & 1.314$^{{\rm 4}}$ & 81.22 $\pm46.40$ & 7 & 0.48 &
    2619.25 & 4.39 $\times$ 10$^{{\rm 52}}$ \\
    GRB~080319B & 0.937$^{{\rm 5}}$ & 124.86 $\pm3.10$ & 17 & 1.11 &
    2757.78 & 1.98 $\times$ 10$^{{\rm 52}}$ \\
    GRB~080319C & 1.950$^{{\rm 6}}$ & 29.55$\pm9.41$ & 3 & 0.01 &
    538.89 & 2.40 $\times$ 10$^{{\rm 52}}$ \\
    GRB~080605 & 1.640$^{{\rm 7}}$ & 29.55 $\pm9.41$ & 9 & 0.29 &
    2622.20 & 7.63 $\times$ 10$^{{\rm 52}}$ \\
    GRB~090102 & 1.547$^{{\rm 8}}$ & 29.30 $\pm3.23$ & 3 & 0.20 &
    512.07 & 1.29 $\times$ 10$^{{\rm 52}}$ \\
    GRB~090424 & 0.544$^{{\rm 9}}$ & 49.47 $\pm2.27$ & 6 & 0.12 &
    5744.63 & 1.08 $\times$ 10$^{{\rm 52}}$ \\
    GRB~090510 & 0.903$^{{\rm 10}}$ & 5.66 $\pm1.88$ & 1 & 0.52 &
    1761.86 & 1.16 $\times$ 10$^{{\rm 52}}$ \\
    GRB~091020 & 1.710$^{{\rm 11}}$ & 38.92 $\pm4.89$ & 4 & -0.24 &
    354.11 & 1.14 $\times$ 10$^{{\rm 52}}$ \\
    GRB~100814A & 1.440$^{{\rm 12}}$ & 174.72 $\pm9.47$ & 17 & 0.57 &
    482.72 & 1.01 $\times$ 10$^{{\rm 52}}$ \\
    GRB~100906A & 1.727$^{{\rm 13}}$ & 114.34 $\pm1.59$ & 14 & 0.00 &
    636.66 & 2.10 $\times$ 10$^{{\rm 52}}$ \\
    GRB~110213A & 1.460$^{{\rm 14}}$ & 48.00 $\pm16.00$ & 6 & -0.95 &
    498.70 & 1.08 $\times$ 10$^{{\rm 52}}$\\
    GRB~110422A & 1.770$^{{\rm 15}}$ & 25.77 $\pm0.60$ & 7 & 0.03 &
    2167.60 & 7.60 $\times$ 10$^{{\rm 52}}$ \\
    GRB~110503A & 1.613$^{{\rm 16}}$ & 10.05 $\pm3.41$ & 1 & -0.05 &
    2799.08 & 7.81 $\times$ 10$^{{\rm 52}}$ \\
    \hline
  \end{tabular}
\end{table*}

\subsection{BAT data}
\label{sec:batobs}

BAT  data  are  used  in   both  simulation  methods  detailed  in  \S
\ref{sec:modelling}.   The  data,  including  auxiliary data  and  the
Tracking  and   Delay  Relay  Satellite  System   (TDRSS)  data,  were
downloaded  from the  UK \textit{Swift}  Science Data  Centre (UKSSDC;
http://www.swift.ac.uk/swift\_portal/).  These raw data were processed
using the standard \textit{Swift} routine {\sc batgrbproduct}.\par

\begin{figure*}
  \begin{center}
    \includegraphics[angle=0]{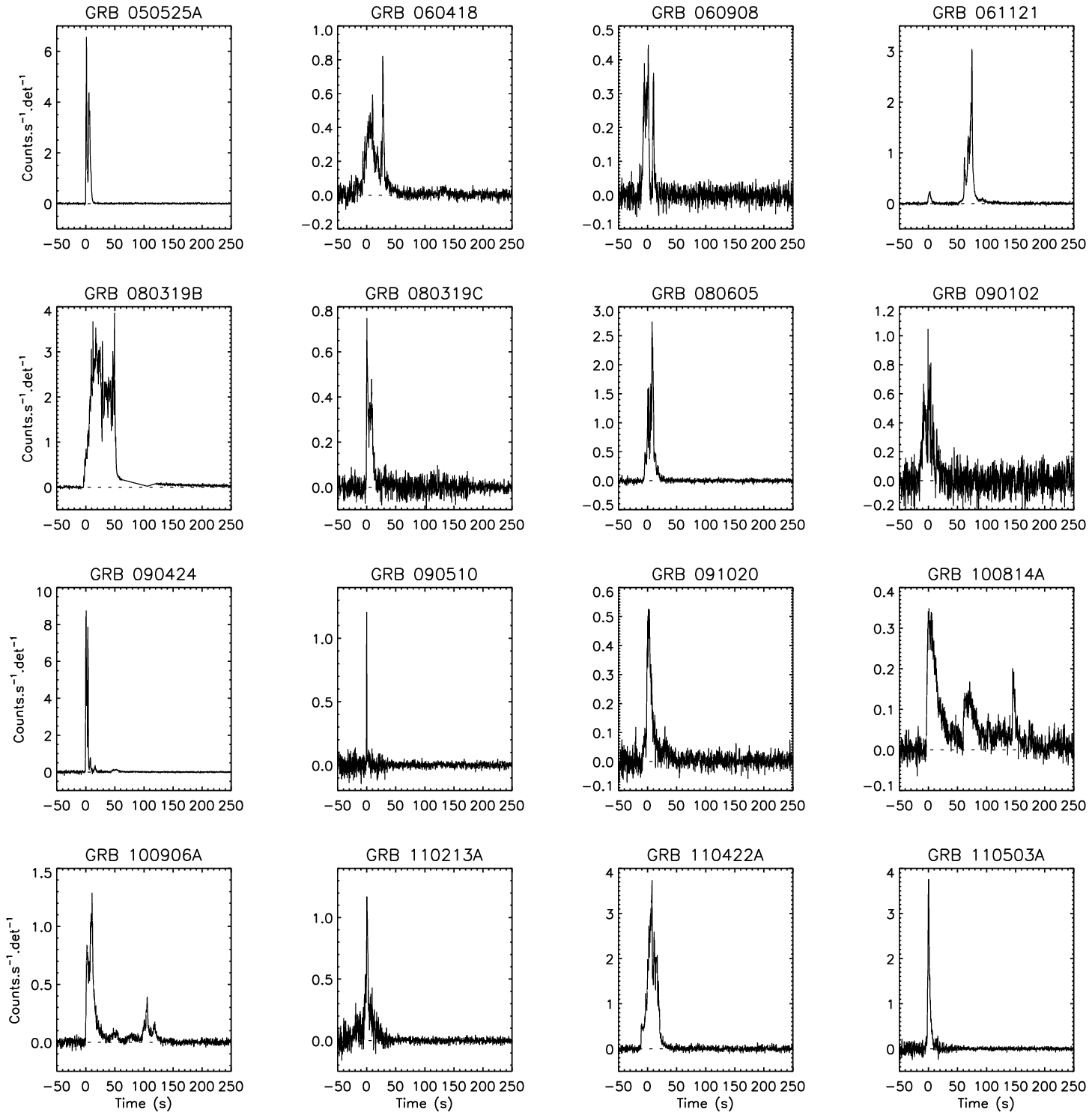}
  \end{center}
  \caption{High-energy  15--350  keV  BAT  light  curves  of  the  low
    redshift bright burst sample.  The  light curves are binned at 256
    ms to allow clear  identification of structure within each.  There
    is also a dashed line in each panel to show a zero count rate.}
  \label{obslc}
\end{figure*}

One of the resulting outputs of {\sc batgrbproduct} is a series of GRB
light  curves taken  at multiple  temporal resolutions  in  either one
total (15--350  keV) band or  the four individual  bands (15--25\,keV,
25--50\,keV, 50--100\,keV  and 100--350\,keV)  as used in  typical BAT
analysis.   The observed  BAT light  curves for  the  bright ($z_{{\rm
    orig}}<2$) sample are shown in Figure \ref{obslc}, where the light
curves shown  are those  obtained when the  four bands are  summed for
each burst.\par

\subsection{XRT data}
\label{sec:xrtobs}

\textit{Swift}  X-Ray  Telescope (XRT;  \citealt{2005SSRv..120..165B})
data have not been simulated in this work, however, as discussed in \S
\ref{sec:pulsim} below,  the second method used to  simulate BAT light
curves does involve  modelling of both BAT and  XRT data following the
method   of  \citet{2010MNRAS.403.1296W}.   Therefore   an  additional
criterion for inclusion in  the sample is that \textit{Swift} executed
a prompt  slew to the  burst, thus providing the  necessary early-time
X-ray observations, although we emphasise that there is no requirement
that the  X-ray data overlap the  period of activity  detected by BAT.
The speed at which  \textit{Swift} slews its narrow--field instruments
to a new GRB is determined solely by observing constraints; including,
for  example, the  proximity of  the  source to  the Sun  or Moon,  or
whether such a  source is occulted by the  Earth.  Thus this selection
criterion does not bias our sample.\par

\section{Modelling}
\label{sec:modelling}

Two different  methods of simulation  were developed.  The  first uses
the observed BAT data in the spectral range that would fall within the
15--25\,keV band if  the burst in question had  occur at the simulated
higher redshift ($z_{{\rm sim}}$).   The second method uses the prompt
pulse model detailed in  \citet{2010MNRAS.403.1296W}, as fitted to the
observed BAT and XRT data, evolving the characteristic times, energies
and normalizing fluxes  that describe each pulse. We  describe each in
the following sections (with further details in the appendices).\par

In  both  cases,  creation  of  the simulated  light  curves  required
accounting  for   the  effects  of  cosmological   time  dilation;  of
band-shifting,  as individual  photons are  redshifted  (this included
consideration  of the  BAT response  and  changes in  background as  a
function of energy); and of declining flux due to increased luminosity
distance as defined in Equation \ref{eq:lf}.\par

\begin{equation}
  F(E_1\mapsto E_2) = \frac{L(E_1(1+z)\mapsto E_2(1+z))}{4 \pi (D_{L}(z))^{2} }.
  \label{eq:lf}
\end{equation}

Here  $F(E_1\mapsto  E_2)$ is  the  flux of  the  bin  in a  specified
observing   band   between    photon   energies   $E_1$   and   $E_2$,
$L(E_1(1+z)\mapsto E_2(1+z))$  is the corresponding  luminosity in the
same  {\em  de-redshifted}  band  and  $D_{L}(z)$  is  the  luminosity
distance  of  the source  at  redshift  $z$.  To calculate  luminosity
distances we adopted a standard $\Lambda CDM$ cosmology with $H_{0}$ =
71 km\,s$^{{\rm  -1}}$\,Mpc$^{{\rm -1}}$, $\Omega_{m}$  $=$ $0.27$ and
$\Omega_{\Lambda}$ $=$ 0.73.\par

\subsection{Method 1: High-redshift simulations made directly from BAT data}
\label{sec:batmethod}

In  our  first method  (Method  1 hereafter)  we  aimed  to produce  a
simulated 15--25\,keV observer-frame light  curve for a GRB located at
$z_{\rm sim}$. To do so, we  extracted a light curve from the observed
BAT data for a burst at  $z_{\rm orig}$ at 64\,ms resolution using the
standard {\sc batbinevt} routine, which also subtracts the background.
Specifically, we produced the  light curve in the 15$\zfac$--25$\zfac$
\,keV range, where for convenience we have defined $\zfac$ as shown in
Equation \ref{z_fac1}.\par

\begin{equation}
  \zfac=\frac{1+z_{\rm sim}}{1+z_{\rm orig}}.
  \label{z_fac1}
\end{equation}

So, in an instance where $z_{{\rm orig}} = 1$ and $z_{{\rm sim}} = 3$,
$\zfac=2$ and we were required  to use {\sc batbinevt} to extract both
a 15--25 keV and 30--50 keV light curve.\par

\begin{figure}
  \begin{center}
    \includegraphics[width=9.2cm,angle=0]{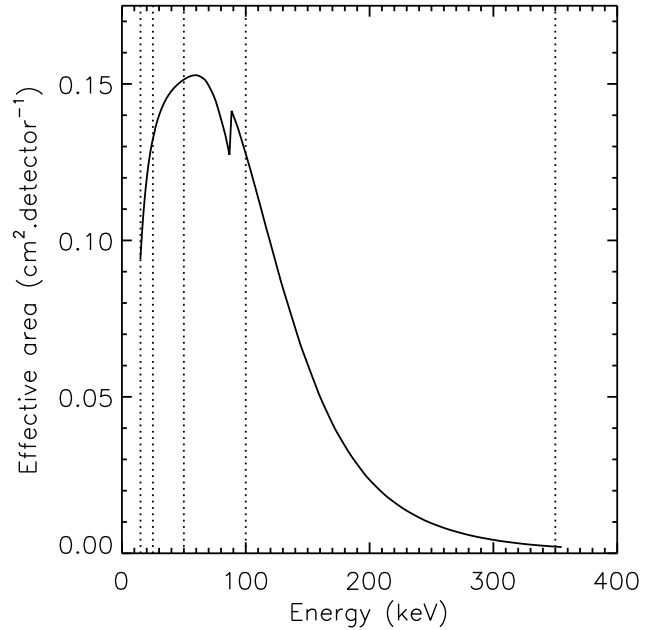}
  \end{center}
  \caption{The  BAT  effective area  per  detector  as  a function  of
    energy. The  vertical dotted lines  show the energy ranges  of the
    four standard BAT channels as described is \S \ref{sec:batobs}.}
  \label{fig:aeff}
\end{figure}

We needed to account for the energy dependence of the BAT sensitivity,
$A(E)$, which is  shown in Figure \ref{fig:aeff}. As  can be seen, for
higher energies the integrated effective area is small. For simplicity
we assumed  that the  effective area of  the detector over  the energy
band of  interest could be taken  simply as an average  of $A(E)$ over
that band (strictly speaking one  should weight by the spectral energy
distribution  of the  source, but  this will  usually not  be  a large
correction, and we neglect it at this stage so as to avoid introducing
any model fitting).  Hence, for  each bin in the original light curve,
defined by  start and  end times  $T_1$ and $T_2$,  we can  relate the
photon counts in the simulation to the original counts via:\par

\begin{equation}
  \begin{array}{lll}
    C_{\rm sim}(T_1\zfac\mapsto T_2\zfac)&=  &\zfac^2
    \left(\frac{D_{\rm L,orig}}{D_{\rm L,sim}}\right)^{2}\frac{\int\limits_{15}^{25}A\left( E \right).dE}{\int\limits_{15\zfac}^{25\zfac}A\left(E\right).dE} \\
    && \times C_{\rm orig}(T_1\mapsto T_2),\\
    
    & =&  QC_{\rm orig}(T_1\mapsto T_2).
    \end{array}
  \label{eq:ctocsimp}
\end{equation}

Thus the quantity $Q$ provides the factor by which the observed counts
in  the 15$\zfac$--25$\zfac$  light curve  at $z_{\rm  orig}$  must be
reduced to calculate the 15--25\,keV light curve at $z_{\rm sim}$. $Q$
is  dependent  on  three   contributions:  the  change  in  luminosity
distance, the redshifting of the photons and the change in sensitivity
of the BAT instrument between  the two energy bands.  Providing $Q<1$,
we can  therefore produce a new  simulated light curve  with the right
noise characteristics  by simply losing  an appropriate number  of the
original counts from each bin,  prior to rebinning back to an observed
64\,ms time resolution.\par

Finally, we  ensured that the  background noise was adjusted  to match
that expected  in the  15--25\,keV band, as  measured in  the original
data.  These steps are described in more detail in Appendix~A.\par

\subsection{Method 2: Simulations using the prompt pulse model} 
\label{sec:pulsim}

The disadvantage in  using the BAT data directly, as  we did in Method
1,  is  the  restriction   imposed  by  the  finite  energy  response.
Consequently,  we  required  that   25$\zfac$  $<$  350  keV  for  the
simulation  to  be within  the  BAT  coverage.  Aside from  this,  the
uncertainties due to  the shape of the BAT  spectral response increase
significantly  at  the  upper  end  of  this  energy  range,  reducing
confidence in  the simulations conducted at the  highest redshifts. In
practice, then, we are limited to $\zfac\lesssim 10$.\par

An  alternative and more  flexible approach  is to  fit models  to the
temporal and spectral behaviour of the real GRBs, and use these models
as   the   basis    for   simulations.   Specifically,   we   followed
\citet{2010MNRAS.403.1296W}  who fit  both  the observed  BAT and  XRT
light curves simultaneously using a  combination of the pulse model of
\citet{2009MNRAS.399.1328G}, first proposed to explain the steep decay
phase  (SDP)   as  a   result  of  high   latitude  emission,   and  a
phenomenological  afterglow   model  \citep{2007ApJ...662.1093W}.  The
\citet{2009MNRAS.399.1328G} model includes  spectral evolution, and so
being able to fit the SDP, which  is not seen in the BAT light curves,
helps  constrain the  decay of  all pulses.  This is  the  reason that
availability of  early-time XRT data was  one of the  criteria used to
select our sample (see \S \ref{sec:xrtobs}).\par

Using  this methodology (Method  2 hereafter)  we calculated  the flux
light curves of all four primary BAT and the two XRT channels (the BAT
channels  being as  described in  \S 3.2  and the  XRT  channels being
0.3--15 keV  and 1.5--10 keV),  allowing a more extensive  analysis of
the  high-energy characteristics  of the  prompt emission.  To conduct
such simulations  the parameters  of each prompt  pulse, and  also the
X-ray  afterglow, needed to  be evolved  to the  values that  would be
observed by placing the burst progenitor at a higher redshift.  Figure
1  of  \citet{2010MNRAS.403.1296W}, which  is  a  representation of  a
single  pulse   temporal  profile,  shows   the  characteristic  times
associated with each pulse.  These include the time over which a pulse
rises ($T_{\rm r}$), the time  at which the pulse peaks ($T_{\rm pk}$)
and  the arrival  times of  the first  and last  photons  ($T_{0}$ and
$T_{\rm  f}$).  The  values of  these at  the simulated  redshift were
calculated  using time  dilation as  shown in  Equation \ref{eq_chrt},
where  $T_{\rm sim}$  and  $T_{\rm  orig}$ correspond  to  any of  the
characteristic  timescales  at the  simulated  and observed  redshifts
respectively.\par

\begin{equation}
  T_{\rm sim}= \zfac\times T_{\rm orig}.
  \label{eq_chrt}
\end{equation}

In this model,  the spectrum of each of the pulses  is a time evolving
exponentially  cut--off  power-law.   The  peak energy  of  the  $\nu
F_{\nu}$ spectrum is  defined at the time when the  pulse peaks in the
light curve of  the GRB, $T_{\rm pk}$. When a pulse  is simulated at a
higher  redshift, time dilation  causes $T_{\rm  pk}$ to  occur later.
Furthermore, the  photon energy at  the peak of the  spectrum, $E_{\rm
  pk,sim}$,  reduces as  a  pulse  is placed  at  a higher  luminosity
distance.\par

\begin{equation}
  E_{\rm pk,sim}= \frac{E_{\rm pk,orig}}{\zfac}.
  \label{eq_che}
\end{equation}

Finally, each  pulse has a  normalizing flux, $S_{\rm  pk,sim}$, which
was  calculated from  the original  normalisation,  $S_{\rm pk,orig}$,
using Equation \ref{eq_chn}.\par

\begin{equation}
  S_{\rm pk,sim} = \left( \frac{K_{\rm orig}}{K_{\rm sim}} \right)
  \left( \frac{D_{\rm L,orig}}{D_{\rm L,sim}} \right)^{2} S_{\rm pk,orig}.
  \label{eq_chn}
\end{equation}

As previously  discussed, the k--correction,  $K$, is used  to convert
the fluence  in either  the observed or  simulated energy band  to the
bolometric fluence of the pulse.\par

Once the pulse parameters were adjusted, we ran the models through the
same  software used to  produce light  curves such  as those  shown in
\citet{2010MNRAS.403.1296W}. This  provided smooth 64  ms light curves
that were background  subtracted in each of the  four BAT bands. These
model light  curves formally contain no statistical  noise from either
the source or  the background, so finally we  added artificial scatter
to account for both, as elaborated in Appendix B.\par

\subsection{Rate triggering algorithm}
\label{sec:ratealg}

To provide a fair comparison  to the observed high-redshift GRBs it is
essential to  establish whether or  not a given simulated  light curve
would  actually  have triggered  \textit{Swift}/BAT.   This requires  an
algorithm that emulates the BAT  trigger process. In practice, that is
very hard to do fully  realistically, so we restrict ourselves to only
considering ``rate" triggers\footnote{``Rate"  triggers occur when the
  count-rate  over all  the  detectors exceeds  some given  threshold.
  Being a coded-mask instrument, the data from BAT can also be used to
  reconstruct  sky  images  allowing  variable  point  sources  to  be
  identified  as  ``image"-triggers.    The  latter  are  particularly
  valuable  for  finding  long-lasting,  low  peak-luminosity  events,
  although in  practice the large  majority of bursts are  detected as
  rate triggers.}.\par

For each rate-trigger  we considered two durations measured  by BAT: a
trigger and background duration.  Of those background regions measured
by BAT,  we took that which occurs  prior to  the region of  the light
curve  during which the  burst is  active. The  trigger duration  is a
narrower window  that constantly runs  through the on-board  BAT data.
The total fluence within this window is cumulated and if the signal to
noise ratio of  the total fluence exceeds a  predefined threshold, the
algorithm  reports  a  trigger  to  begin  normal  \textit{Swift}  GRB
observations.\par

\begin{figure}
\begin{center}
\includegraphics[angle=0,width=9.2cm]{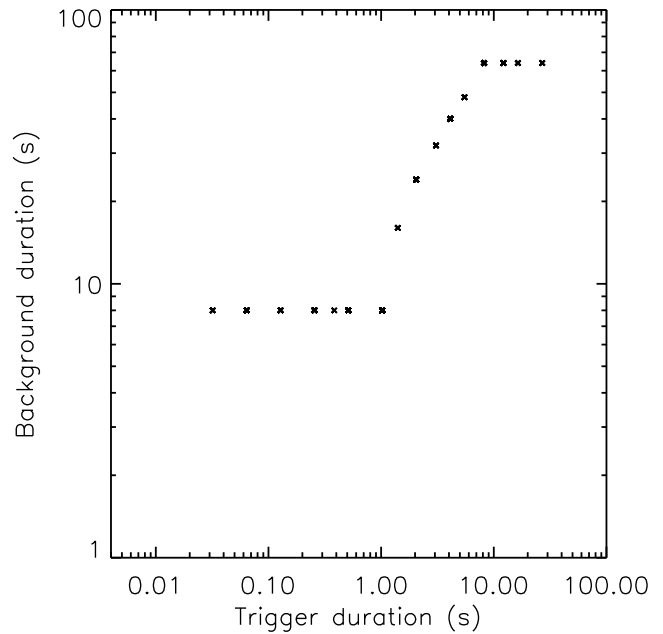}
\end{center}
\caption{Background duration as a function of trigger duration used in
  the  BAT  detection algorithm  for  all  bursts  that triggered  the
  \textit{Swift}/BAT (excluding GRB~041228, as discussed). }
\label{fig:trigtimes}
\end{figure}

To define the  trigger durations used in our  algorithm, we considered
all \textit{Swift}  bursts which triggered BAT. For each  burst we  looked at  the durations  of the
trigger and the background, as well  as the energy ranges used to find
the  standard  combinations  for  typical BAT  triggers.   These  real
combinations of  trigger and background durations are  shown in Figure
\ref{fig:trigtimes}.\par

The BAT algorithm has a well defined set of combinations of background
and trigger durations. We  exclude GRB~041228 (with a trigger duration
of 64  ms and a  background duration of  40 seconds) due to  having an
unusual trigger,  not successfully used  in any subsequent  burst. The
majority  of  bursts  trigger  on  a  1024 ms  trigger  and  8  second
background duration, in the 25  -- 100 keV range.  This corresponds to
channels 2 and 3 of BAT.\par

We selected a background region that occurred before the trigger time,
which  was chosen  to maximise  the total  fluence within  the trigger
duration.  Using these  regions, we applied Equation \ref{eq:fentrig},
taken  from  \citet{2003AIPC..662..491F},  to  calculate  the  trigger
significance.\par

\begin{equation}
  S = \frac{\left( C_{j} - 2^{j-10} B \right)^{2}}{2^{j-10} B + \sigma_{min}^{2}}.
  \label{eq:fentrig}
\end{equation}

In Equation \ref{eq:fentrig}, $C_{j}$ is the total number of counts in
the region of  the light curve containing the  most counts, whilst $B$
is the number of background counts expected in 1024 ms. $j$ is defined
by the binning  of the light curve, with each bin  being $2^{j}$ ms in
duration. In effect, the numerator of Equation \ref{eq:fentrig} is the
background   corrected   number   of   counts   within   the   trigger
duration. $\sigma_{min}$  ensures there is  always a minimum  value of
variance.  However,  as   we  specified  a  realistic,  non-negligible
background, we  did not  include $\sigma_{min}$ when  calculating $S$.
It was $S$ that we used  to judge whether the detectability of a given
light curve using BAT.\par

Only the combinations shown in Figure \ref{fig:trigtimes} (discounting
that  for  GRB~041228) were  used  to  define  background and  trigger
times. Additionally, to further mimic \textit{Swift} and also minimise
processing  time, the  triggers were  ordered  by how  often they  had
occurred  within  the  real  burst  population.  Once  a  trigger  was
successfully satisfied  at 6.4$\sigma$ the  algorithm stopped checking
further triggers.\par

It  is  important  to  note  that  our  rate-triggering  algorithm  is
conservative,  in  the  sense   that,  while  it  may  designate  some
simulations ``undetected"  when the real  BAT would have  triggered on
them (since the {\em Swift} in-flight software includes more elaborate
criteria,  such as  image triggers),  it  is unlikely  that any  light
curves we claim  are ``detected" would have been  missed by BAT.  Thus
our simulations  should provide a  fair comparison to the  {\em Swift}
high-$z$ sample,  which all would  have triggered BAT based  upon this
algorithm.\par

\subsection{Measuring $T_{90}$}
\label{sec:t90alg}

The  simulated light curves  produced files  in the  standard template
used by  the routine {\sc  battblocks} for calculating  $T_{90}$. {\sc
  battblocks}  uses  a  Bayesian  analysis  to  find  robust  duration
measures  \citep{1998ApJ...504..405S}  including  the  time  intervals
$T_{90}$,  $T_{50}$ and, if  required, any  duration specified  by the
user.\par

Aside  from  modifying  the  values  of  flux,  time  and  error,  key
parameters within the data files, such as the start, stop and exposure
times  of the significant  proportions of  the light  curve had  to be
altered.   These were  adjusted  using the  temporal  factor shown  in
Equation \ref{eq_chrt}.\par

We ran {\sc  battblocks} at a temporal resolution  of 1024 ms. Binning
the simulated light curves at  such a resolution, rather than directly
from the 64 ms output, proved  useful, as when the signal became weak,
the coarser temporal binning  allowed for more reliable calculation of
$T_{90}$.\par

\section{A detailed case study of 16 bright bursts}
\label{sec:disc}

Taking the 16 bursts identified  with bright pulses, we simulated each
at a variety  of redshifts in the range  2 $\leqslant$ $z$ $\leqslant$
12. The simulations were conducted  at small increments in redshift of
$\Delta z$ $=$ $0.125$ up to $z$ $=$ 5, and in steps of $\Delta z$ $=$
$0.5$ thereafter.\par

Both simulation methods used  these values for $z_{\rm sim}$, however,
Method  1 was restricted  in some  cases due  to the  limited spectral
coverage  from BAT.   In  these  cases the  simulations  could not  be
performed  past $\zfac$  $=$ 6.  This  affected four  of the  bright
bursts: GRB~050525A,  GRB~080319B, GRB~090424 and  GRB~090510, although
in all cases the triggering algorithm failed to detect structure prior
to the cut off in the simulations.\par

In practice,  at higher redshifts  the measured durations  become very
uncertain,  so  for  each  burst   we  ran  100  simulations  at  each
redshift. The measured durations  (where the burst still satisfied the
trigger criteria) were then averaged, using the geometric mean.\par

\subsection{Results of Method 1}
\label{sec:meth1res}

\begin{figure*}
  \begin{center}
    \includegraphics[angle=0]{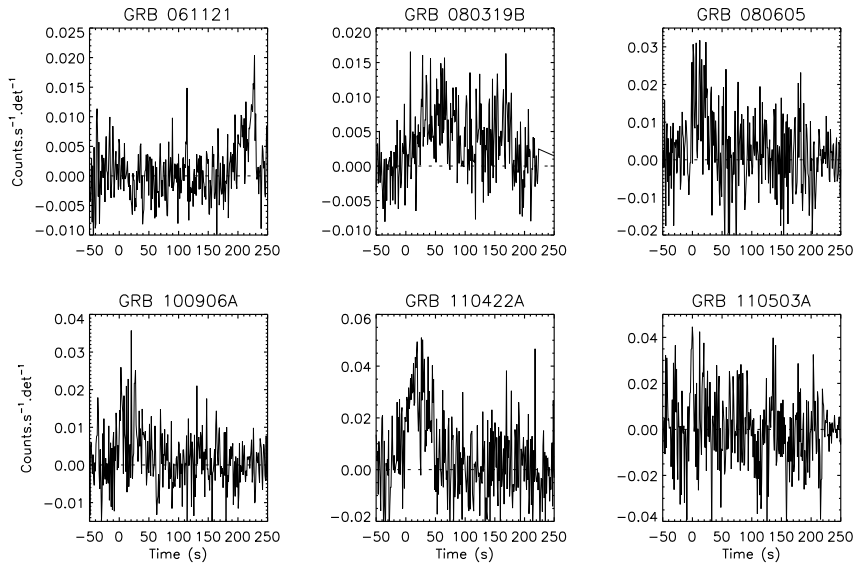}
  \end{center}
  \caption{15--25 keV BAT light curves of the bright burst subset when
    simulated at $z_{\rm sim}$ $=$ 6 using Method 1.  The light curves
    are binned at 1024 ms to allow clearer identification of structure
    within each.  Only those bursts  which were detected using the BAT
    algorithm at $z_{\rm sim}$ $=$ 6 are shown.  There is also  a dashed
    line in each panel to show a zero count rate.}
  \label{fig:lcm1}
\end{figure*}

Using Method 1  to simulate the 15--25 keV light  curves of the bright
subset revealed  that only  six bursts would  be detected in  at least
90\% of the  simulations at $z_{\rm sim}$ $=$  6. Example light curves
at this simulated redshift are shown in Figure \ref{fig:lcm1}.\par

\begin{figure*}
  \centering
  \includegraphics[angle=0,width=8.3cm]{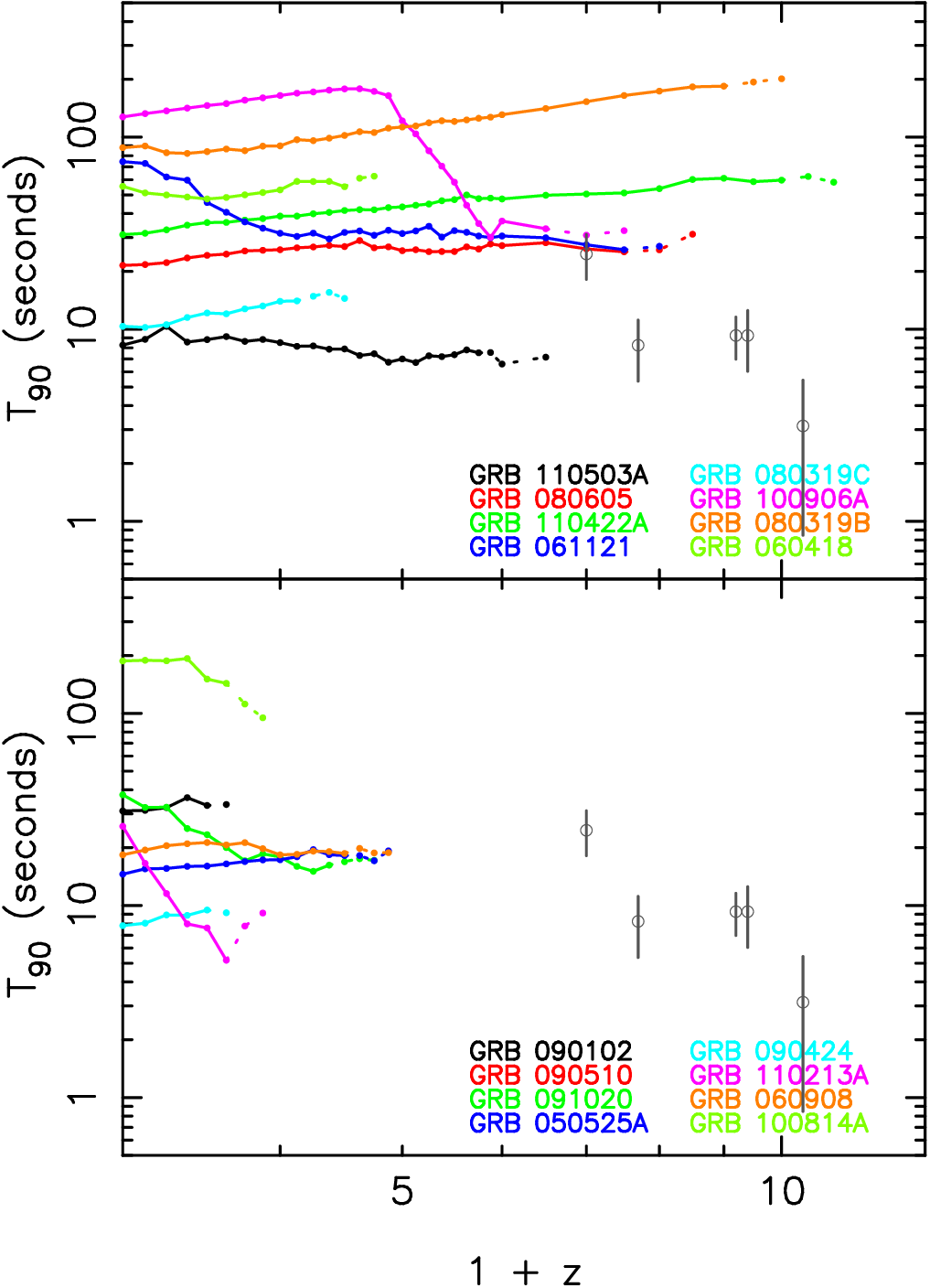}
  \quad
  \includegraphics[angle=0,width=8.3cm]{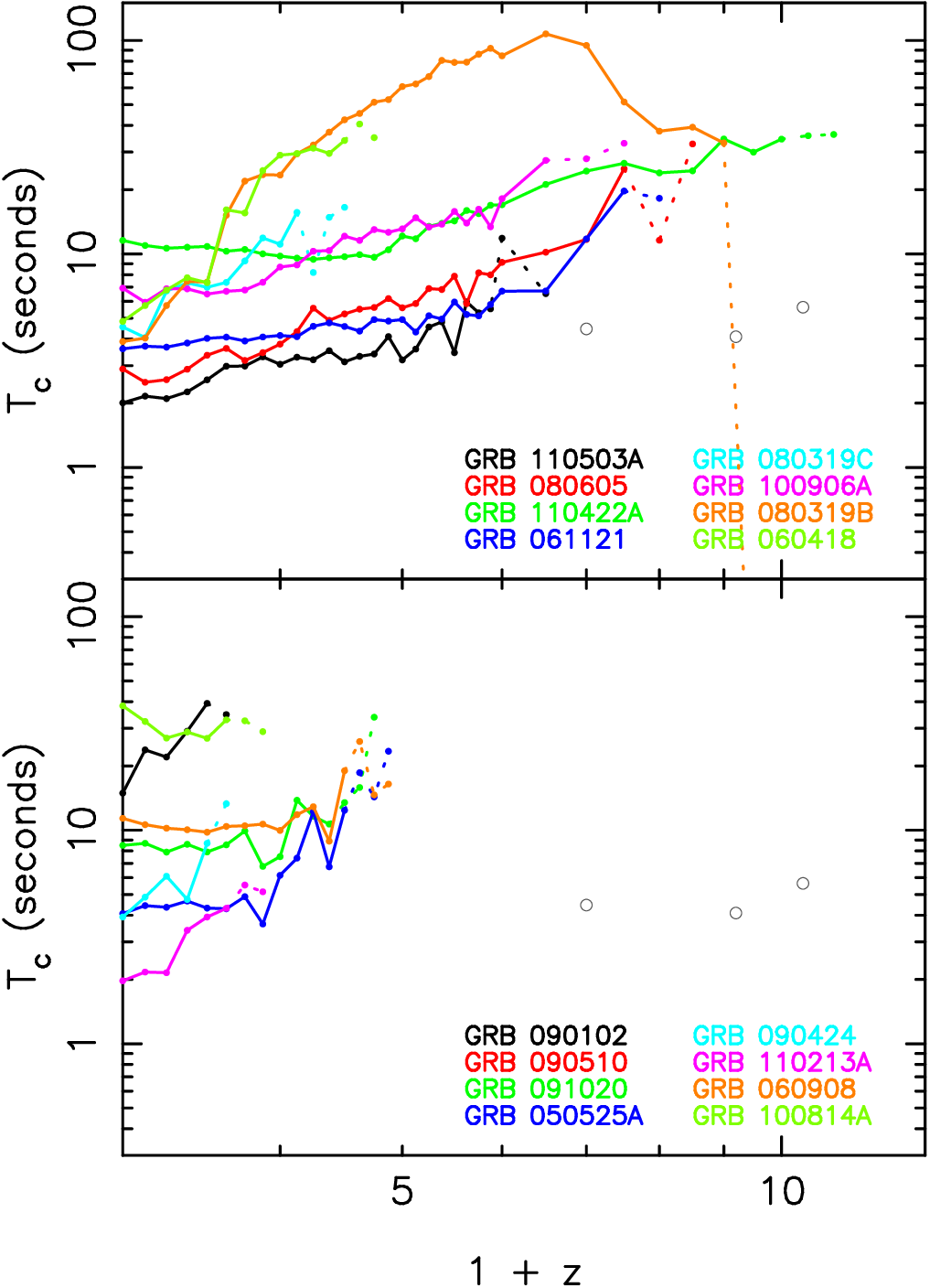}
  \caption{Predictions of observed  durations using Method 1 ($T_{90}$
    and $T_{c}$ on the left  and right, respectively) of bright bursts
    as  they would  have  been seen  if  they had  occurred at  higher
    redshift.   Each coloured  line  corresponds to  simulations of  a
    single GRB in the bright sample, with each duration being averaged
    over 100 repeated simulations. The sample has been split according
    to the luminosity of the  brightest pulse in each burst (see Table
    \ref{tab_bright}), with the most luminous 8 being shown in the top
    panel. Solid lines show where  the burst was detected in a minimum
    90\% of the repeated simulations,  the dotted lines show where the
    burst  was detected in  less than  90\% but  greater than  50\% of
    these repeats.  The grey open circles correspond to the 15--25 keV
    $T_{90}$ and  $T_{c}$ values of the  observed high-redshift bursts
    in the left and right panels respectively.}
  \label{fig:evodurm1}
\end{figure*}

The standard {\sc battblocks} algorithm was run on all simulated light
curves. The BAT  triggering algorithm was then used  to identify which
of the light  curves contained detectable structure in  the 15--25 keV
range. The  geometric average of  the triggered light curves  was then
taken at  each $z_{\rm sim}$ for  each burst.  These are  shown in the
left  panels  of  Figure  \ref{fig:evodurm1},  which  also  shows  the
measured $T_{90}$ for the observed high-redshift bursts.\par

In  Figures \ref{fig:evodurm1}  and  \ref{fig:evodurm2} the  simulated
bursts  have  been  divided  into  two subsets  based  on  their  peak
luminosities  to make it  easier to  see the  evolution of  each.  The
eight  bursts with  the brightest  peak  luminosities are  in the  top
panel, with the remaining eight shown in the bottom panel.\par

The same  process was repeated for  $T_{c}$.  The results  of this are
shown in the right panels of Figure \ref{fig:evodurm1}.  Once more the
same  analysis has  been conducted  on the  observed 15--25  keV light
curves of  the high-redshift sample  for reference. Only three  of the
six bursts were  bright enough to define a  bright core exclusively in
the 15--25 keV range.\par

\subsection{Results of Method 2}
\label{sec:meth2res}

\begin{figure*}
  \begin{center}
    \includegraphics[angle=0]{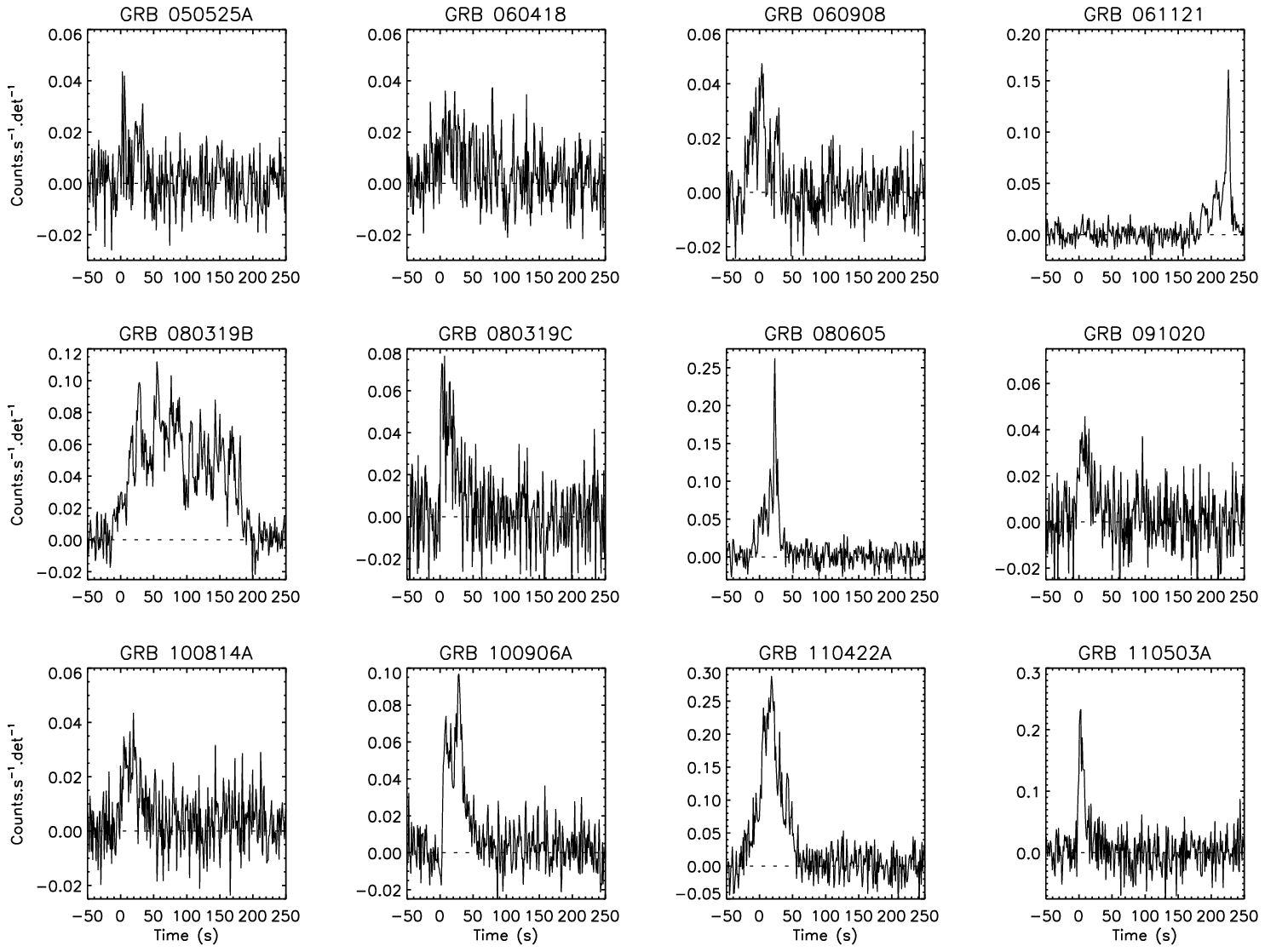}
  \end{center}
  \caption{15--350  keV BAT light  curves of  the bright  burst subset
    when simulated at  $z_{\rm sim}$ $=$ 6 using  Method 2.  The light
    curves are  binned at 1024  ms to allow clearer  identification of
    structure  within each.   Only  those bursts  which were  detected
    using the BAT  algorithm at $z_{\rm sim}$ $=$  6 are shown.  There
    is also a dashed line in each panel to show a zero count rate.}
  \label{fig:lcm2}
\end{figure*}

Using Method  2 there were  12 bursts in  the bright sample  that were
detected  when simulated  at  $z_{\rm  sim}$ $=$  6.  Examples of  the
15--350 keV light curves produced for these bursts are shown in Figure
\ref{fig:lcm2}, where  the light curves  are binned at 1024 ms  to make
the presence of structure easier to discern by eye.\par

\begin{figure*}
  \centering
  \includegraphics[angle=0,width=8.3cm]{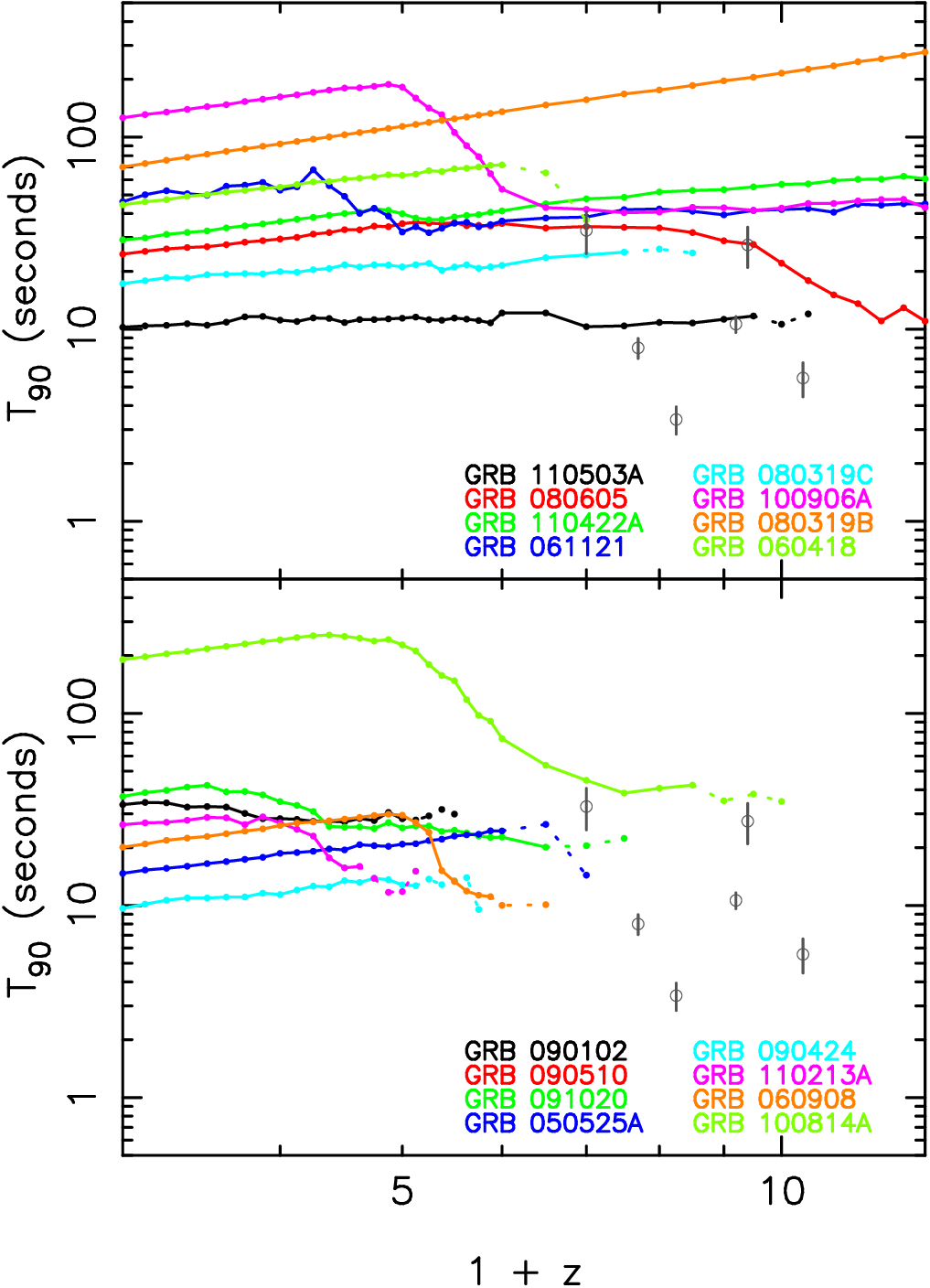}
  \quad
  \includegraphics[angle=0,width=8.3cm]{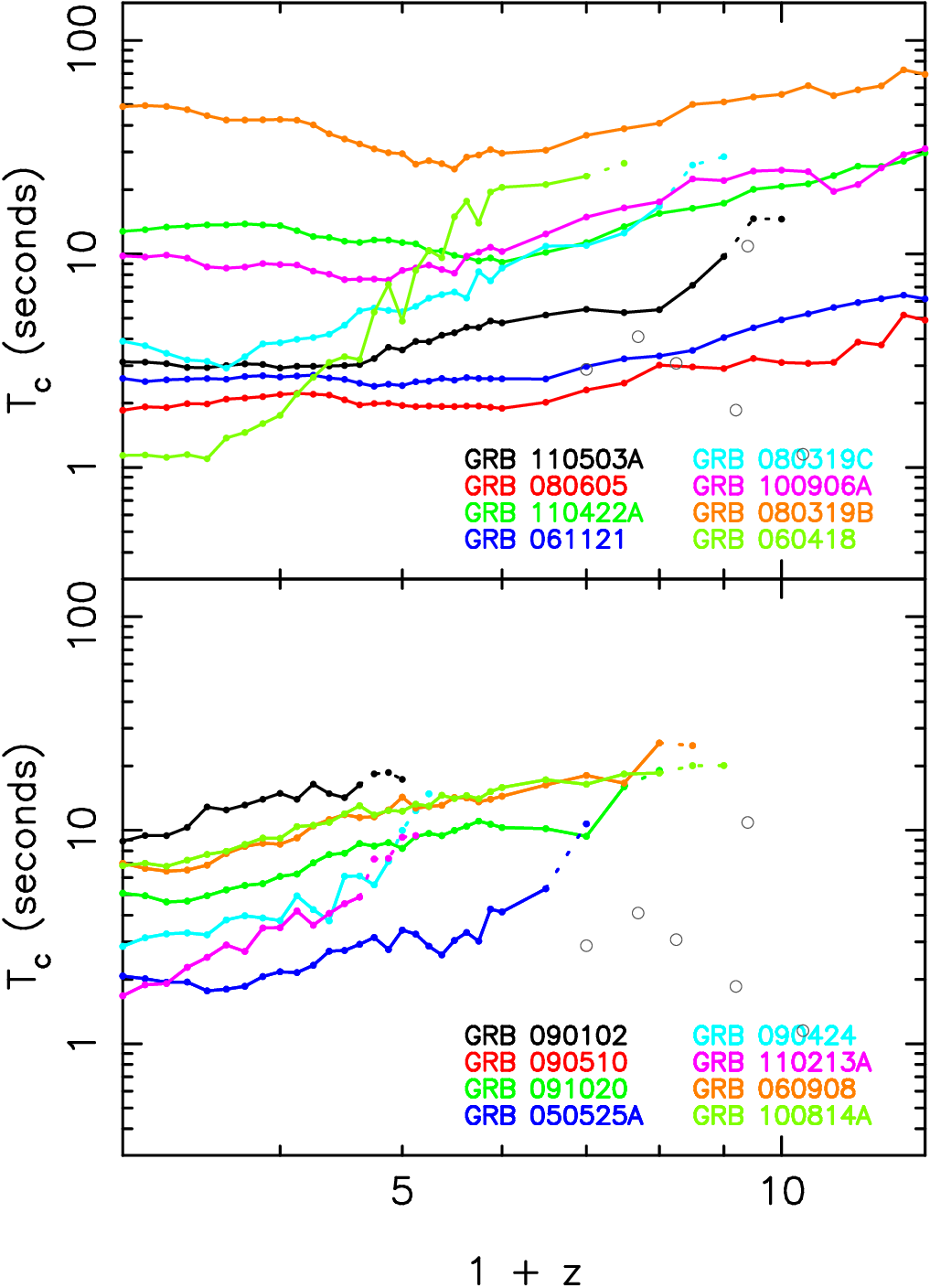}
  \caption{Predictions of observed  durations using Method 2 ($T_{90}$
    and $T_{c}$ on the left  and right, respectively) of bright bursts
    as  they would  have  been seen  if  they had  occurred at  higher
    redshift.   Each coloured  line  corresponds to  simulations of  a
    single GRB in the bright sample, with each duration being averaged
    over 100 repeated simulations. The sample has been split according
    to the luminosity of the  brightest pulse in each burst (see Table
    \ref{tab_bright}), with the most luminous 8 being shown in the top
    panel. Solid lines show where  the burst was detected in a minimum
    90\% of the repeated simulations,  the dotted lines show where the
    burst  was detected in  less than  90\% but  greater than  50\% of
    these repeats.   The grey open  circles correspond to  the 15--350
    keV  $T_{90}$ and  $T_{c}$  values of  the observed  high-redshift
    bursts in the left and right panels respectively..}
  \label{fig:evodurm2}
\end{figure*}

We also  plotted the evolution of  $T_{90}$ with redshift  for all the
bright  bursts.   This   is  shown  in  the  left   panels  of  Figure
\ref{fig:evodurm2},  where the  four channel  15--350  keV simulations
obtained using Method 2 were  summed to create a single channel.  This
15  -350 keV light  curve was  analysed using  {\sc battblocks}  in an
identical manner  to both Method 1 and  the regular \textit{Swift}/BAT
analysis. Once more the  observed $T_{90}$ values of the high-redshift
sample   are    also   shown   in   the   left    panels   of   Figure
\ref{fig:evodurm2}.\par

We derived $T_{c}$  for all the simulated light  curves produced using
Method 2. The  averages of the light curves  with structure that would
trigger   BAT   are   shown    in   the   right   panels   of   Figure
\ref{fig:evodurm2}.   Method 2  allowed us  to find  $T_{c}$  over the
entire  15--350 keV  range.   All six  of  the observed  high-redshift
sample were bright enough to define a bright core across this spectral
range.\par

\subsection{Evolution of $T_{90}$}
\label{sec:evot90}

\begin{figure}
  \begin{center}
    \includegraphics[width=6cm,angle=270]{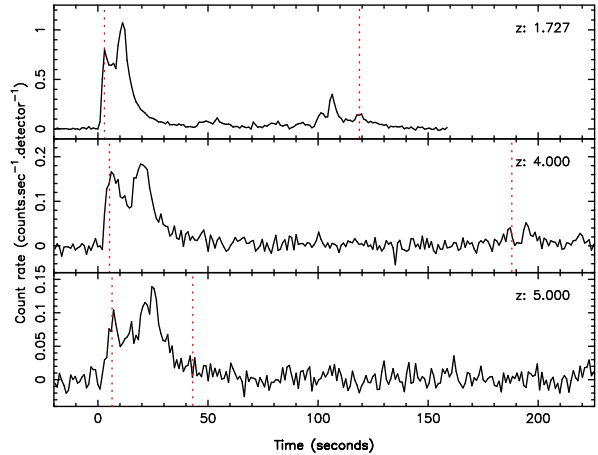}
  \end{center}
  \caption{15  -- 350 keV  light curve  simulations of  GRB~100906A at
    three redshifts.  From top to  bottom these redshifts are  $z$ $=$
    1.727, 4,  and 5. $z$  $=$ 1.727 is  the observed redshift  of the
    burst. The  binning of this  light curve is  1024 ms and  the red
    dotted lines  show the  start and end  of the  identified $T_{90}$
    period.  Note that the vertical  scaling is different in the three
    panels.   The figure  illustrates  how the  observed duration  can
    decrease rapidly with redshift when some phase of the emission, in
    this case the second major peak, drops below detectability.}
  \label{simb_100906A}
\end{figure}

The  behaviour  of each  burst  in  the  left-hand panels  of  Figures
\ref{fig:evodurm1}  and  \ref{fig:evodurm2}  can  be  reasonably  well
understood by  considering the  effects of cosmological  time dilation
leading   to  an  increase   in  duration   (seen  most   clearly  for
GRB\,080319B),  together  with  the  opposing tendency  for  declining
signal-to-noise  and band-shifting  to reduced  duration.   The latter
effects depend on the morphology of the light curve: in particular the
fact that GRB prompt emission often shows hard-to-soft evolution means
that later peaks are likely  to become undetected before earlier peaks
resulting in the conspicuous decline in observed duration seen in some
cases (e.g., GRB~100906A  beyond $z_{\rm sim}$ $=$ 5;  see also Figure
\ref{simb_100906A}).    In  some  cases,   such  as   GRB~110503A  and
GRB~080319C, the  duration is almost  independent of redshift,  due to
the  loss  of  detected flux  in  the  wings  of the  prompt  emission
approximately cancelling out the effects of time dilation.\par

As  can  be  seen  in  Figure  \ref{simb_100906A},  as  $z_{\rm  sim}$
increases the  signal to  noise ratio of  structure within  the light
curve  significantly  reduces.   As  such,  it is  expected  that  the
uncertainty   in  the   average  values   obtained  for   $T_{90}$  at
high-redshift will be high. These uncertainties have not been included
on Figures  \ref{fig:evodurm1} and  \ref{fig:evodurm2} as it  was felt
this would further crowd each panel.\par

\begin{figure}
  \begin{center}
    \includegraphics[angle=0,width=8.3cm,clip]{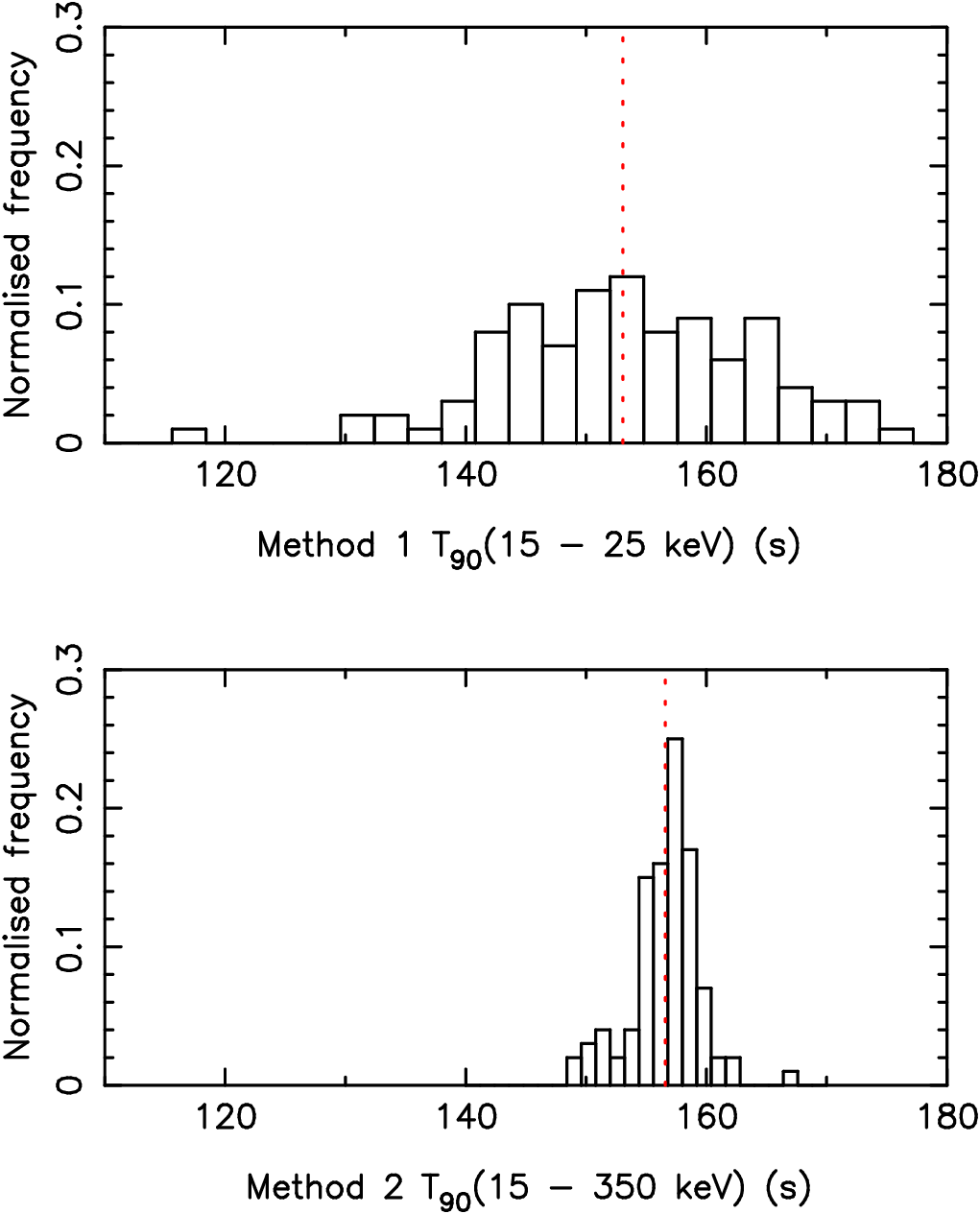}
  \end{center}
  \caption{Histograms showing  the scatter in  $T_{90}$ estimates from
    repeated simulations at $z_{\rm sim}$ $=$ 6 for GRB~080319B, using
    both Method 1 (top panel)  and Method 2 (bottom panel). Using both
    methods,  all  100  repeats   yielded  detections  using  the  BAT
    triggering  algorithm.   The red  dotted  line  shows the  average
    T$_{90}$ found in both cases.}
  \label{fig:t90scat}
\end{figure}

Instead  to  give  an  estimate  of the  scatter,  we  include  Figure
\ref{fig:t90scat}, which  shows histograms of the  values obtained for
all 100  repeated simulations of  GRB~080319B at $z_{\rm sim}$  $=$ 6.
Both  methods are  shown, with  Method  1 yielding  a greater  scatter
(Method 1: 153.1  seconds, Method 2: 156.6 seconds).  This is expected
as Method 1 only provides data  on a single BAT channel, and therefore
has fewer  counts in  each light  curve. As such  the signal  to noise
ratio is  poorer when  compared to a  corresponding simulation  of the
same burst using Method 2.\par

The   net    result   shown   in    Figures   \ref{fig:evodurm1}   and
\ref{fig:evodurm2} is  that the bulk  of the simulations  lie between
$15\,{\rm s}<T_{90}<50\,{\rm  s}$ at both $z=2$  and $z=6$, consistent
with   the  mild   evolution  in   $T_{90}$  seen   in   the  observed
\textit{Swift}  sample with  redshifts (Figure  \ref{t90_vs_z}).  Note
that at higher redshifts  the reduced signal-to-noise ratio means that
bursts may become undetected in  some simulations.  In the left panels
of   Figures  \ref{fig:evodurm1}   and   \ref{fig:evodurm2}  this   is
illustrated  by the  lines  being  plotted as  solid  while the  burst
remained detected  in 90\%  of cases, and  dotted for  detection rates
between  50\% and 90\%.   Not surprisingly,  this increased  noise can
lead to  somewhat erratic evolution  in the measured $T_{90}$  in some
cases near the detection limit.\par

\subsection{Evolution of $T_{c}$}
\label{sec:evotc}

The evolution of  $T_{\rm c}$ with redshift is  shown in the right-hand
panels of Figures  \ref{fig:evodurm1} and \ref{fig:evodurm2}.  As with
$T_{90}$, the evolution can be  tracked for a broader range in $z_{\rm
  sim}$ using Method  2, due to the improved signal  to noise ratio of
each  light  curve. In  most  instances $T_{\rm  c}$  can  be seen  to
increase  even  for  bursts  where $T_{90}$  reduces  with  increasing
redshift. In some cases this rate of increase is significantly greater
than would naively be expected due to time dilation: this can arise if
the brightest  peak happens  to have a  relatively soft  spectrum, and
hence declines more  rapidly than the majority of  the light curve due
to band shifting as we simulate the burst at higher redshift.\par

Overall  there are  fewer  instances  of a  decline  in duration  with
redshift than  was the case  for $T_{90}$, supporting  the proposition
that  $T_{\rm c}$  is less  sensitive  to the  loss of  pulses in  the
background noise.\par

Finally, the  evolution with  $z_{\rm sim}$ is  smoother for  Method 2
than Method 1. This is again largely due to the improved statistics of
higher count rates.\par

\section{Comparison with the known high-redshift population}
\label{sec:compks}

\begin{figure*}
  \begin{center}
    \includegraphics[angle=0]{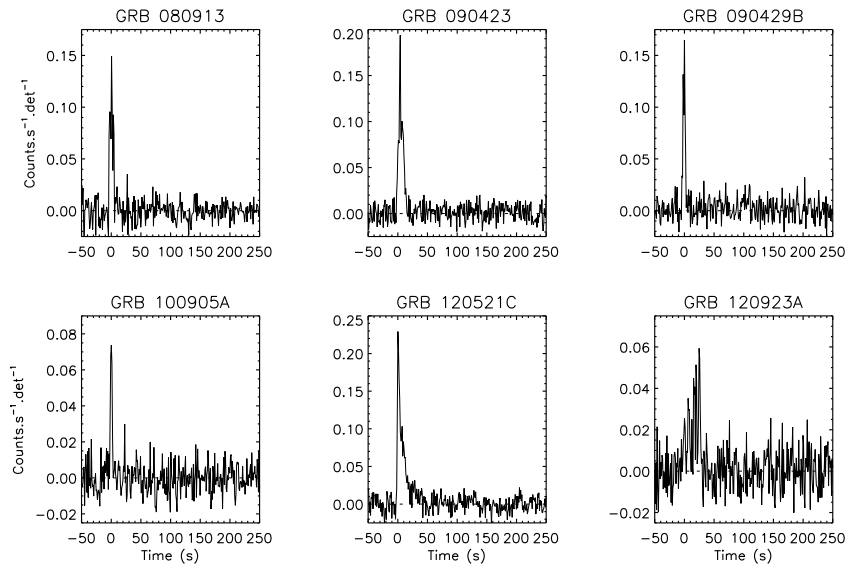}
  \end{center}
  \caption{15--350 keV  BAT light curves of  the high-redshift subset.
    The  light  curves  are  binned   at  1024  ms  to  allow  clearer
    identification of  structure within each.  There is also  a dashed
    line in each panel to show the zero count rate.}
  \label{fig:lchiz}
\end{figure*}

Figure \ref{fig:lchiz}  shows the observed  light curves of  the known
high-redshift population. Using  the triggering algorithm described in
\S \ref{sec:ratealg} we verified that all six would have triggered BAT
in  both  the  15--25 keV  and  15--350  keV  range.  As  our  trigger
algorithm only accounts for  ``rate triggers'', and does not implement
an ``image  trigger'' search, we are being  conservative in comparison
to BAT.   It is likely  that for a  small range in  simulated redshift
shortly after a  burst becomes too faint to  cause a ``rate trigger'',
``image  triggers'' would  be detected  therefore recovering  a higher
fraction of  the simulated  burst population. The  measured durations,
$T_{90}$ and $T_{c}$, obtained  for the high-redshift burst sample are
included in Table \ref{tab:hizbursts}. It is important to note at this
point that we have not included GRB~050904, which was identified as an
``image trigger'' by BAT. We verified using our trigger algorithm that
the  BAT  light  curve  could  not have  reached  a  ``rate  trigger''
threshold   using   any   combination   of  background   and   trigger
durations.\par

\begin{table*}
  \centering
  \caption{The high-redshift GRB subset.  Values of $T_{90}$ are those
    obtained  from the  standard  BAT analysis  in  the energy  ranges
    stated. $T_{c}$  was obtained using the algorithm  described in \S
    \ref{sec:deftc}. Missing entries in  this Table are due to failure
    of  the software  to identify  necessary bright  structure  in the
    light  curves.  (Redshifts  from: 1.  \citep{2009ApJ...693.1610G},
    2.  \citep{2009Natur.461.1254T},  3.  \citep{2011ApJ...736....7C},
    4. \citep{2012GCN..13348...1T}, 5. \citep{2012GCN..13802...1L})}
  \label{tab:hizbursts}
  \begin{tabular}{@{}cccccc}
    \hline
    \hline
    GRB & Redshift & $T_{90}(15-25~{\rm keV})$ & $T_{90}(15-350~{\rm keV})$ 
    & $T_{c}(15-25~{\rm keV})$ &  $T_{c}(15-25~{\rm keV})$\\
    & & (s) & (s) & (s) & (s) \\
    \hline
    GRB~080913 & 6.695$^{{\rm 1}}$ & 8.26 $\pm$2.90 & 8.00 $\pm$0.95 
    & ... & 4.10 \\
    GRB~090423 & 8.2$^{{\rm 2}}$ & 9.28 $\pm$2.29 & 10.62 $\pm$1.00 & 4.10
    & 1.86 \\
    GRB~090429B & 9.4$^{{\rm 3}}$ & 3.14 $\pm$2.29 & 5.57 $\pm$1.12 & 1.15
    & 5.563 \\
    GRB~100905A & $\sim$7.25 & ... & 3.39 $\pm$7.96 & ... & 3.07 \\
    GRB~120521C & $\sim$6$^{{\rm 4}}$ & 24.64 $\pm$6.48 & 32.70 $\pm$7.96 &
    4.42 & 2.88 \\
    GRB~120923A & $\sim$8.4$^{{\rm 5}}$ & 9.28 $\pm$3.24 & 27.46 $\pm$6.49 
    & ... & 10.88 \\
    \hline
  \end{tabular}
\end{table*}

Should time dilation be considered  to be the only mechanism affecting
the observed  high-redshift GRB sample, the durations  of these bursts
when moved to  the local Universe would be  considerably shorter. If a
na\"{i}ve factor  of $1+z$ is  applied, it can  be shown that  all six
would have a rest frame duration of $T_{90}< 5$ seconds.\par

We have already shown, however,  that the evolution of duration is the
result  of  several  effects,  not  just time  dilation.  It  is  also
important to note that one short GRB was included in the bright subset
of  bursts considered  in  detail.  GRB~090510 may  only  be a  single
example,  however  it  conforms  with  observational  trends  and  the
expectation  that  short bursts  are  not  sufficiently  bright to  be
detected  at   even  moderate   redshifts.  This  suggests   that  the
high-redshift subset  are unlikely  to be short  GRBs which  have been
misidentified due to time dilation of their prompt durations.\par

The prompt light curves of  high-redshift GRBs are often considered to
be   the  ``tip  of   the  iceberg''   with  weaker   structure  being
undetectable. This makes it  impossible to simulate high-redshift GRBs
by  blueshifting them into  local Universe  where this  faint emission
would become  visible. To affect a successful  comparison between GRBs
at high- and low-redshifts, we therefore did the reverse: we simulated
observed low-redshift GRBs at high-redshifts.\par

We  took the average  observed redshift  of the  high-redshift sample,
$\bar{z}_{\rm   high}  =7.66$   (this  excludes   the   image  trigger
GRB~050904,  which  would  reduce  the value  to  $\bar{z}_{\rm  high}
=7.46$).  We  then  imposed  an  upper limit  in  redshift  for  those
low-redshift  bursts we  would simulate  at $\bar{z}_{\rm  high}$.  We
included this upper limit ($z_{\rm  orig}< 4$) to mitigate the effects
of  any   redshift  dependent   change  in  observed   light  curve
morphology due to evolution in the progenitor population.\par

We  then  simulated all  114  bursts  within  the pulse-fitted  sample
meeting  our  redshift  criterion   ($z_{\rm  orig}<  4$)  at  $z_{\rm
  sim}=\bar{z}_{\rm high}$.   Using the  same process adopted  for the
bright  subset, we  repeated the  simulation of  each burst  100 times
using Methods 1 and 2.  We checked which of the simulated light curves
for  the  114  bursts would  have  caused  a  BAT trigger,  using  our
triggering algorithm.  Those light  curves which were bright enough to
trigger BAT at $z_{\rm  sim}=\bar{z}_{\rm high}$ were then analysed to
find both $T_{90}$ and $T_{c}$.   For each simulated burst we averaged
the values of  $T_{90}$ and $T_{c}$ over the  repeats which garnered a
detection.   These averages were  then used  in the  statistical tests
outlined below.  It  is important to note that  only those bursts with
detections in  a minimum  of 50\% of  their repeated  simulations were
retained.\par

To  quantify  whether the  observed  high-redshift population  differs
significantly  from simulations of  low-redshift bursts  at comparable
redshifts  we performed K-S  tests to  compare the  measured durations
described in this work.\par

\begin{figure*}
  \begin{center}
    \includegraphics[width=5.6cm,angle=0,clip]{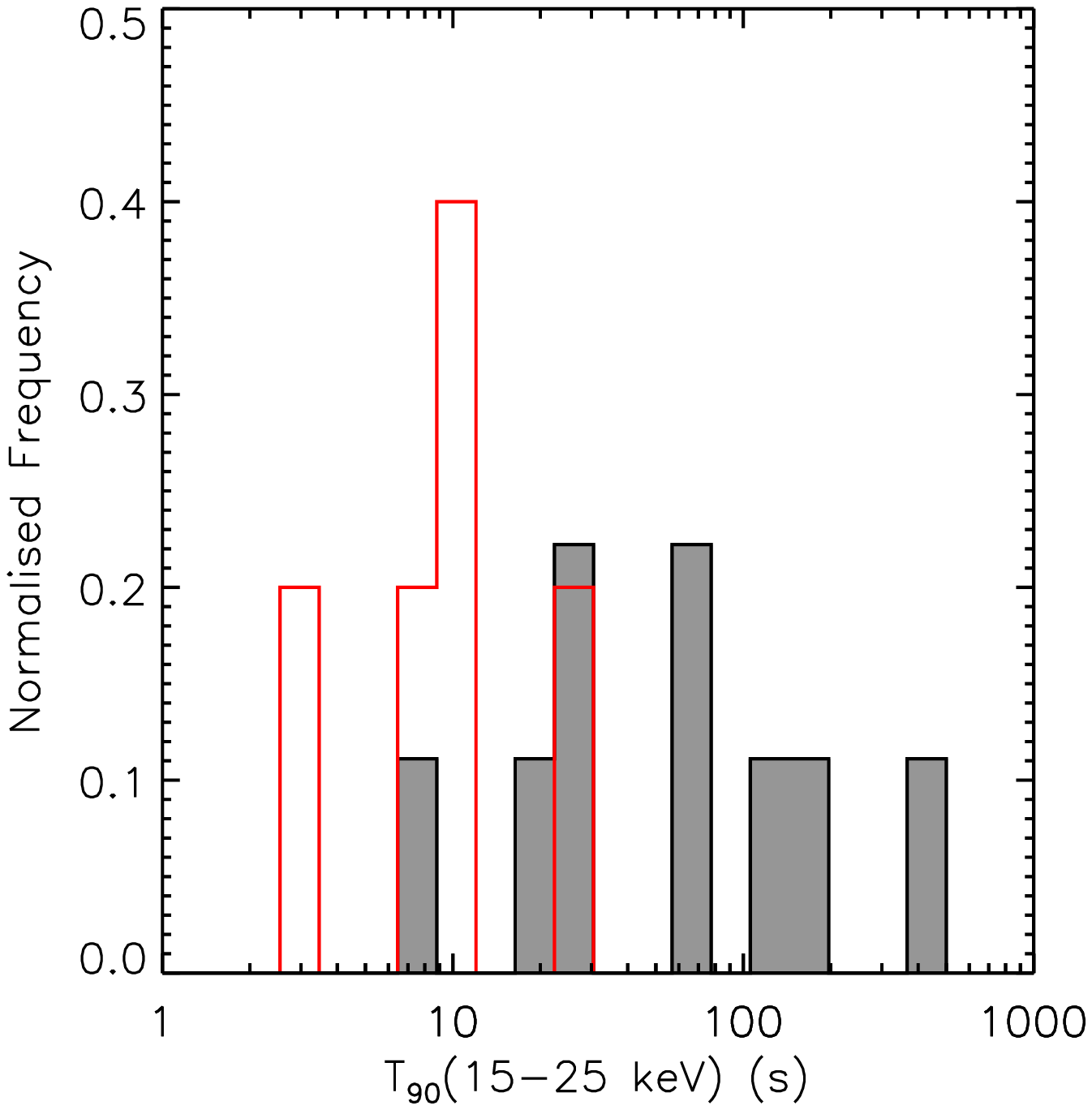} \quad
    \includegraphics[width=5.6cm,angle=0,clip]{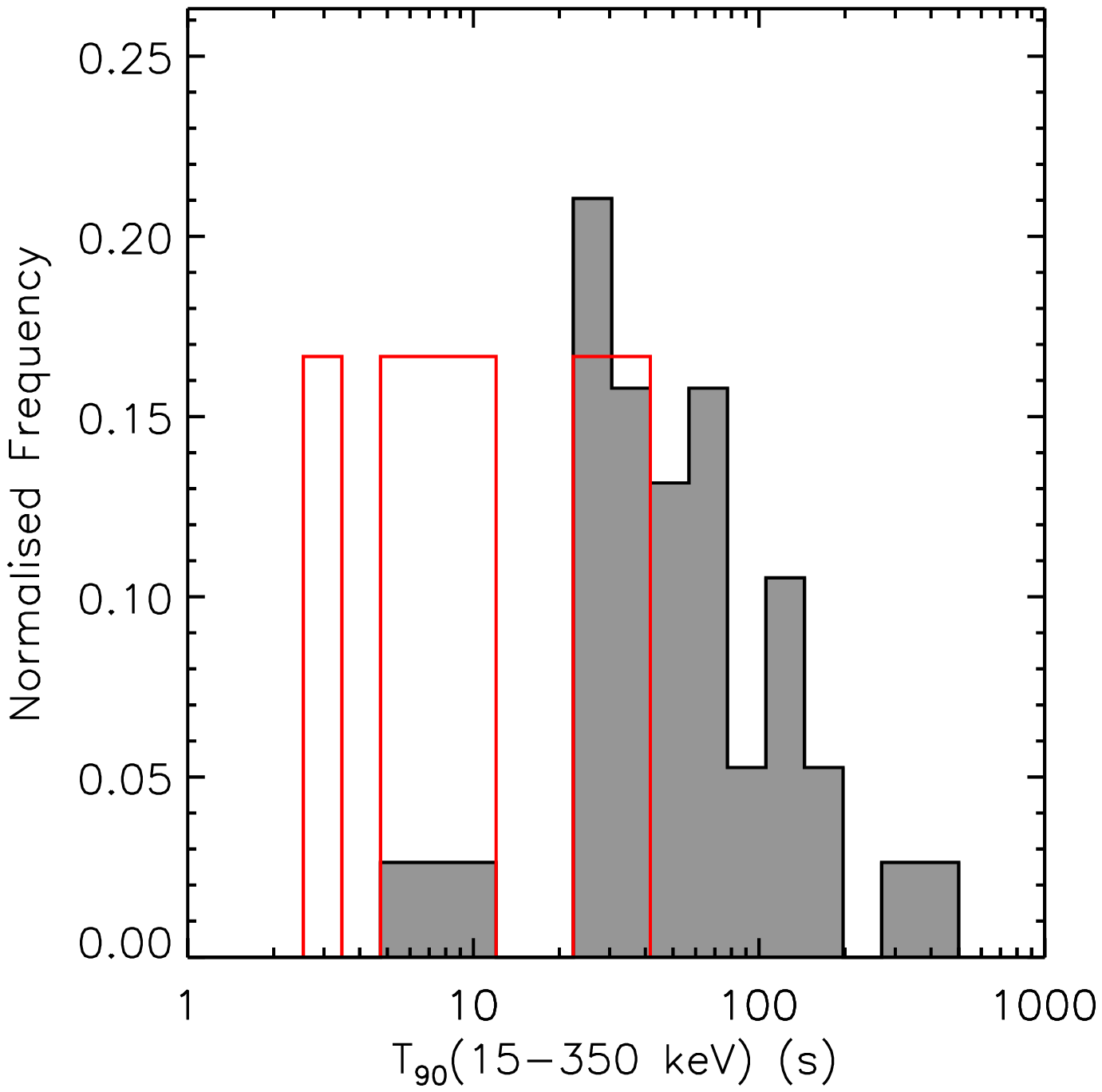} \quad
    \includegraphics[width=5.6cm,angle=0,clip]{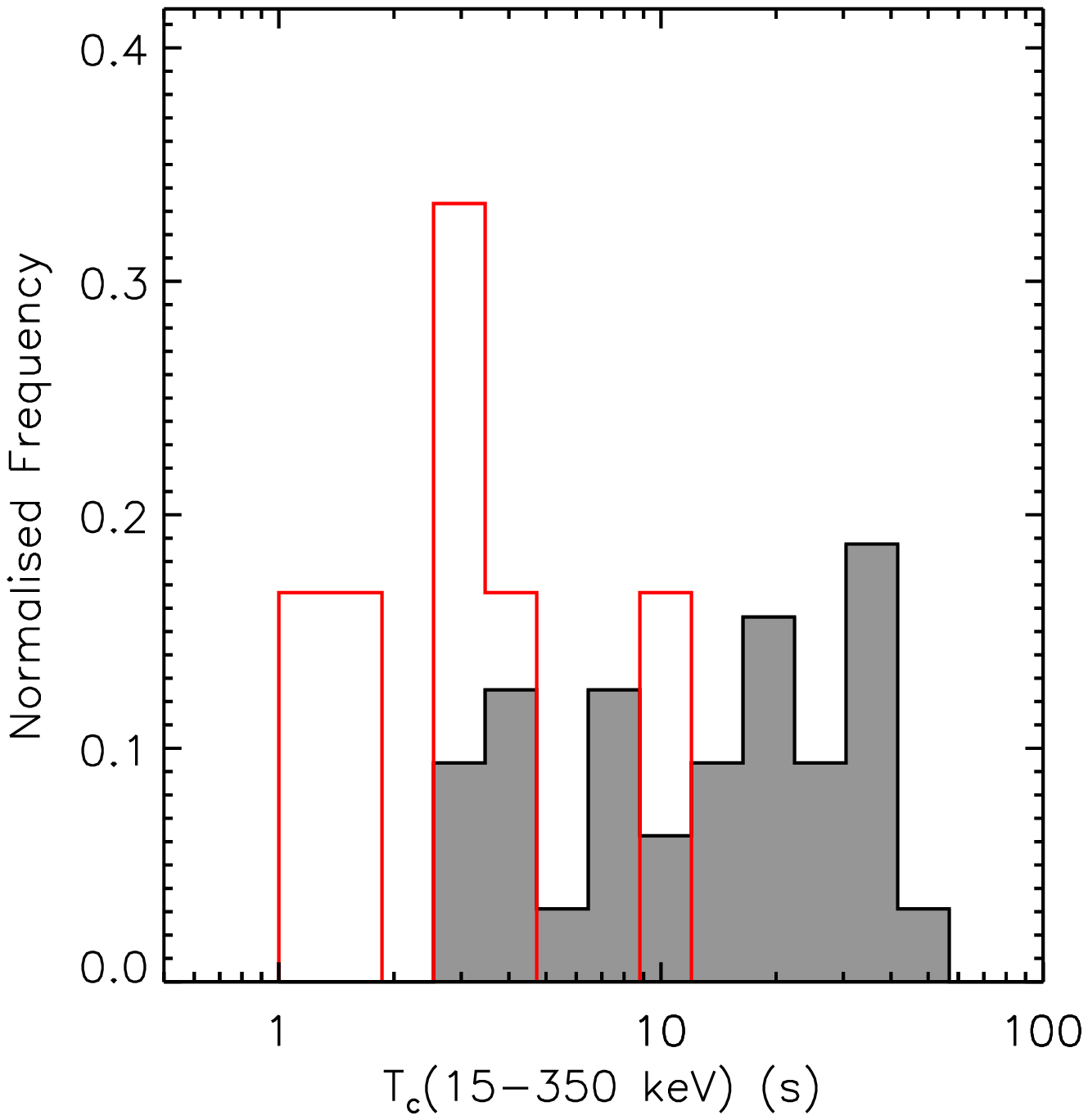} \\
  \end{center}
\caption{Histograms showing  the normalised duration  distributions of
  low-redshifts GRBs  simulated at $\bar{z}_{\rm  high}$ (filled grey)
  and   the   observed    high-redshift   (open   red)   samples   for
  $T_{90}\left(15-25{\rm   keV}\right)$   measured   from   Method   1
  simulations  (left),  $T_{90}\left(15-350{\rm keV}\right)$  measured
  from  Method  2   simulations  (middle)  and  $T_{c}\left(15-350{\rm
    keV}\right)$  measured  from  Method  2  simulations  (right)  K-S
  tests.}
\label{fig:hists}
\end{figure*}

We were able  to perform a K-S  test on three of the  four measures of
duration.   We  were   unable  to  compare  the  $T_{c}\left(15-25{\rm
  ~keV}\right)$ values obtained  via Method 1 as only  three bursts in
the high-redshift sample had sufficient  signal to noise in the 15--25
keV band  to yield a  $T_{c}$ value. The  results for these  tests are
reported as  the top four  lines of Table  \ref{tab:ksresults}, whilst
histograms  corresponding to the  three successfully  tested durations
are shown in Figure \ref{fig:hists}.\par

\begin{table*}
  \centering
  \caption{Results from K-S  tests. The Method used (1  or 2) is shown
    in the first column. The  second shows the duration being compared
    and the  corresponding spectral range over which  it was measured.
    $z_{\rm orig}$  describes the range of  observed low-redshift GRBs
    that  were simulated  at  $z_{\rm sim}$  for  comparison with  the
    high-redshift sample.  For the  last four tests,  comparisons were
    being made  to a moderate  redshift sample, for which  the $N_{2}$
    population spanned  $3 <  z_{\rm high} <  4$. $N_{1}$  denotes the
    number of  bursts in the simulated  sample with a  minimum of 50\%
    detections at $\bar{z}_{\rm high}$  and successful output from the
    associated duration  algorithm. $N_{2}$ corresponds  to the number
    of observed  high-redshift bursts with  a measurable value  of the
    relevant duration.   $D$ is the K-S statistic,  which measures the
    maximum  distance  between  the  cumulated  distributions  of  the
    samples being compared, and  $p$ is the resulting probability that
    the two samples arise from  a common population. Rows in which the
    probability  and maximum distance  are missing  are due  to either
    $N_{1}$ or $N_{2}$ being too small to allow a K-S test.}
 \label{tab:ksresults}
  \begin{tabular}{@{}cccccccc}
    \hline
    \hline
    Method & Duration & $z_{\rm orig}$ & $z_{\rm sim}$ & $N_{1}$ & $N_{2}$ &
    $D$ & $p$ \\
    \hline
    1 & $T_{90}\left( 15-25~{\rm keV} \right)$ & $<4$ & 7.66 & 9 & 5 & 0.689 
    & 4.95 $\times$ $10^{-2}$ \\
    1 & $T_{c}\left( 15-25~{\rm keV} \right)$ & $<4$ & 7.66 & 8 & 3 & ... 
    & ... \\
    2 & $T_{90}\left( 15-350~{\rm keV} \right)$ & $<4$ & 7.66 & 38 & 6 & 0.649 
    & 1.30 $\times$ $10^{-2}$ \\
    2 & $T_{c}\left( 15-350~{\rm keV} \right)$ & $<4$ & 7.66 & 32 & 6 & 0.677 
    & 9.45 $\times$ $10^{-3}$ \\
    \hline
    1 & $T_{90}\left( 15-25~{\rm keV} \right)$ & $<2$ & 7.66 & 2 & 5 & ... 
    & ... \\
    1 & $T_{c}\left( 15-25~{\rm keV} \right)$ & $<2$ & 7.66 & 1 & 3 & ... 
    & ... \\
    2 & $T_{90}\left( 15-350~{\rm keV} \right)$ & $<2$ & 7.66 & 11 & 6 & 0.667 
    & 3.33 $\times$ $10^{-2}$ \\
    2 & $T_{c}\left( 15-350~{\rm keV} \right)$ & $<2$ & 7.66 & 9 & 6 & 0.667 
    & 4.35 $\times$ $10^{-2}$ \\
    \hline
    1 & $T_{90}\left( 15-25~{\rm keV} \right)$ & $3 < z_{\rm orig} < 4$ & 7.66 
    & 2 & 5 & ... & ... \\
    1 & $T_{c}\left( 15-25~{\rm keV} \right)$ & $3 < z_{\rm orig} < 4$ & 7.66 
    & 3 & 3 & ... & ... \\
    2 & $T_{90}\left( 15-350~{\rm keV} \right)$ & $3 < z_{\rm orig} < 4$ & 7.66 
    & 12 & 6 & 0.667 
    & 2.98 $\times$ $10^{-2}$\\
    2 & $T_{c}\left( 15-350~{\rm keV} \right)$ & $3 < z_{\rm orig} < 4$ & 7.66 
    & 9 & 6 & 0.833 & 5.05 $\times$ $10^{-3}$\\
    \hline
    1 & $T_{90}\left( 15-25~{\rm keV} \right)$ & $<2$ & 3.5 & 13 & 17 & 0.222 
    & 8.11 $\times$ $10^{-1}$ \\
    1 & $T_{c}\left( 15-25~{\rm keV} \right)$ & $<2$ & 3.5 & 9 & 15 & 0.267 
    & 7.49 $\times$ $10^{-1}$ \\
    2 & $T_{90}\left( 15-350~{\rm keV} \right)$ & $<2$ & 3.5 & 31 & 17 & 0.421 
    & 2.81 $\times$ $10^{-2}$ \\
    2 & $T_{c}\left( 15-350~{\rm keV} \right)$ & $<2$ & 3.5 & 10 & 17 & 0.641 
    & 5.62 $\times$ $10^{-3}$ \\
    \hline
\end{tabular}
\end{table*}

The results  of the  K-S tests in  Table \ref{tab:ksresults}  show the
statistics for  comparison between the observed  and simulated samples
are better  when using  Method 2.   This is simply  due to  the larger
spectral range of BAT considered, as  Method 2 allowed us to model all
four of the standard BAT bands. Typically, the count rate for BAT GRBs
peaks in channels 2 and 3,  whilst Method 1 only simulates the softest
channel. As  a result the  signal to noise  ratios for Method  1 light
curves were always poorer than their Method 2 counterparts.\par

Our  null hypothesis,  $H_{0}$,  was that  the  measured durations  of
$T_{90}$ and  $T_{c}$ for both  the observed high-redshift  GRB sample
and simulations  of all bursts  with $z_{\rm orig}<4$ were  drawn from
the same parent population. Table \ref{tab:ksresults} shows low values
of  $p$ for  all  performed  K-S tests,  indicating  it unlikely  that
$H_{0}$ is  true with  significances between 2$\sigma$  and 3$\sigma$.
Our most statistically significant result was for the Method 2 $T_{c}$
K-S test,  where $p=9.45 \times  10^{-3}$, indicating less than  a 1\%
chance that the two samples are  drawn from the same population. A 1\%
probability, for a high tail  only test, corresponds to a 2.33$\sigma$
result. This is well below the standard 3$\sigma$ usually implemented,
which corresponds to a probability of 0.13\%.\par

Our  initial K-S  tests aimed  to compare  GRBs observed  at $z\gtrsim
6$ to the largest  possible sample of low-redshift events simulated
at $z_{\rm sim}$. However, also displayed in Table \ref{tab:ksresults}
are further K-S tests where smaller subsets of the low-redshift sample
are considered. Namely those where $z_{\rm  orig} < 2$ and $3 < z_{\rm
  orig}  < 4$.  In  both cases  the number  of low-redshift  GRBs that
remained detectable  when simulated  at $\bar{z}_{\rm high}$  were too
low to perform  K-S tests on either the  $T_{90}$ or $T_{c}$ durations
as obtained  via Method  1. For Method  2 simulated light  curves, the
additional  K-S  tests  performed  on measured  $T_{90}$  have  chance
probabilities that are approximately  a factor of two less significant
than  the  original  $z_{\rm  orig}<4$  comparison.   These  increased
probabilities are most  likely due to the much  decreased sample size,
$N_{1}$, for both.\par

The  light curves  simulated via  Method  2 have  an increased  chance
probability when  restricting the low-redshift sample  to only $z_{\rm
  orig}<2$. Conversely,  when only using  busts for which $3  < z_{\rm
  orig}  < 4$  the  chance probability  reduces  by a  factor of  two,
showing $p\sim  5 \times  10^{-3}$.  Both these  and the  $T_{90}$ K-S
tests cast  further doubt on any  statistically significant difference
between  the  low-  and  high-redshift  samples, as  none  approach  a
3$\sigma$ significance.\par

To further explore  the evolution of measured durations  as a function
of redshift we also considered  a more moderate change in redshift. We
took all  GRBs from our sample  with $z_{\rm orig} <  2$ and simulated
them at $z_{\rm sim} = 3.5$.  These were then compared to the observed
properties  of all  bursts in  the redshift  range $3  < z  <  4$. The
results  of  the  statistical  tests  for both  $T_{90}$  and  $T_{c}$
measured from  Method 1  and Method  2 light curves  are shown  in the
bottom four lines of  Table \ref{tab:ksresults}. As shown, the results
from the Method 1 light curves have probabilities which firmly show we
cannot reject $H_{0}$.  The results from the Method  2 simulated light
curves  are similar to  the previous  K-S tests,  chance probabilities
$p\sim$2\%    and    $p\sim$0.6\%    for   $T_{90}$    and    $T_{c}$,
respectively.\par

While this draft was in  an advanced state, GRB~130606A was discovered
at    $z=5.91$   \citep{2013ApJ...774...26C}.    The    measured   BAT
$T_{90}\left(   15-350~{\rm  keV}\right)$   $\sim$280s   with  notable
quiescent periods in the $\gamma$-ray  light curve, which is much more
in line with  the expectations of what a  time-dilated burst should be
like (although  we note this  was intrinsically a  particularly bright
event).   Were we  to include  GRB~130606A  in our  high-z sample  the
KS-test for  $T_{90}$ from simulations  produced using Method  2 would
have produced a less significant  probability of 3\% that the observed
and   simulated  high-z   populations   are  from   the  same   parent
distribution. The  corresponding K-S test for $T_{90}$  using Method 1
simulations result in a similarly increased probability of 8\%.\par

Conversely,  as the $T_{90}$  of GRB~130606A  is largely  comprised of
quiescent times,  $T_{c}$ remains  short containing only  fluence from
the  pulse complex  at  approximately 160  seconds  after the  initial
trigger                                              time.\footnote{See
  http://gcn.gsfc.nasa.gov/notices\_s/557589/BA/    for    BAT   light
  curves.} The  K-S test  using Method 2  simulations therefore  has a
reduced  chance  probability  of  0.3\%.  This shows  that  while  the
traditional  $T_{90}$ duration  of GRB~130606A  is longer,  the bright
core  of emission denoting  the time  when the  bursts is  most active
remains  shorter   and  consistent  with  the  other   bursts  in  the
high-redshift sample.\par

\section{Discussion and conclusions}
\label{sec:conc}

We first studied  in detail the simulated evolution of  a sample of 16
\textit{Swift}  GRBs  at  $z<2$   with  high  peak  luminosities.   By
simulating their gamma-ray  light curves as they would  have been seen
if  the  bursts  had  occurred  at  a  range  of  redshifts  ($z_{{\rm
    sim}}>2$), we  have studied the primary  mechanisms responsible in
defining the measured prompt durations of GRBs.\par

We considered two methods of simulation, both of which are detailed in
\S \ref{sec:modelling}.  Method 1 takes  photons received by  BAT, and
considers those that  would fall in the observed  15--25\, keV band as
the burst is  moved to successively higher redshifts.  This method has
the advantage of  being model independent, since it  simply depends on
the detected  rate of  hard photons. On  the other hand,  this reduced
band-width  is   much  less  than   the  full  energy  range   of  BAT
(15--350\,keV) that  is normally used to trigger  and characterise GRB
light curves, and the method is, of course, limited by the high-energy
cut-off of BAT in the maximum  redshift a given GRB can be re-simulated
at.\par
 
Method 2 overcomes these deficiencies by modelling each light
curve    using    the    prompt    pulse   approach    discussed    in
\citet{2010MNRAS.403.1296W}. From these  models, high-energy BAT light
curves can be simulated to  arbitrarily high redshifts using the pulse
temporal  and   spectral  properties  and   typical  background  noise
characteristics.  This  allows for  more  flexible  and wider  ranging
comparisons, although is limited by the validity of the model.\par

With  light curves  in hand,  we first  determined whether  the bursts
would have triggered the BAT ``rate-triggering'' algorithm. Similarly,
in comparing  to the observed  sample of high-redshift bursts  we took
care  to  ensure that  they  also  exceeded  one of  the  rate-trigger
thresholds (in this work ``image trigger'' only bursts were discounted
from  consideration).    If  a  burst  satisfied   the  condition  for
triggering,  we   then  determined   its  duration,  using   both  the
traditional $T_{90}$ and an  alternative measure of the total duration
of bright periods, so-called ``core-time", $T_{c}$.\par

Our  implementation  of  the   BAT  trigger  algorithm  is  not  fully
realistic,  and in  particular some  bursts  may be  found using  more
elaborate BAT  algorithms such as  ``image triggers'', which  would be
considered  non-detections  by  us.   However, because  we  apply  our
analysis consistently  to all  of the bursts  (both known  high-z, and
bright bursts at  low-z), rejecting any from either  subset which fail
our  criteria, that  should  not  introduce any  major  biases in  our
conclusions.\par

In  general  terms,  we  found  that the  measured  durations  of  the
simulated bursts  varied with redshift in  a way that  depended on the
initial structure of the light curve. While cosmological time-dilation
always  works to  lengthen  duration  of the  prompt  emission, it  is
sometimes  countered by the  loss of  some pulses  as they  drop below
detectability,  combined with  the differing  (and  generally shorter)
rest-frame duration  measured in harder  energy bands.  Thus  we might
expect  to  find  comparatively  short  durations  for  the  simulated
high-redshift bursts.  This helps to explain the observed trend in the
population average shown in Figure \ref{t90_vs_z}.\par

One concern  in analyses  of this  sort is that  {\em Swift}  could be
biased  against  finding highly  redshifted  bursts with  time-dilated
light curves due to the restricted periods (typically $<1000$\,s) that
the space-craft dwells in one  location.  This has been highlighted by
the  recent discoveries  of a  class  of ultra-long  GRBs at  moderate
redshifts
\citep{2013ApJ...766...30G,2013arXiv1302.2352L,2011A&A...528A..15G}.
Those studies suggested that some ultra-long bursts ($T_{90}>$~1500 s)
could easily have be missed due  to their emission being spread out in
time.  However,  one important  conclusion from our  work is  that the
normal, bright  LGRB population should not have  observed durations of
more  than a  few  hundred  seconds, even  when  time-dilated at  high
redshift,   suggesting  this   potential  bias   is  unlikely   to  be
significant.\par

Both methods allowed  us to make comparisons with  the observed sample
of very high-redshift \textit{Swift}  GRBs and bursts occurring in the
more  local  Universe.   By   simulating  all  bursts  for  which  the
\citet{2010MNRAS.403.1296W}   pulse-fitting   methodology   had   been
applied, we  considered a sample of 114  low-redshift ($z_{\rm orig}$)
GRBs.  These  were simulated at  $\bar{z}_{\rm high}$ to  allow direct
statistical comparison to the high-redshift subset. The results of the
implemented K-S  do suggest a marginally  significant (99\%) rejection
of  the hypothesis  of no  evolution  of the  GRB population  duration
distribution. Thus  we have shown that the  apparently short durations
of high-\textit{z}  bursts to-date cannot simply be  explained by band
shifting and  sensitivity considerations. On the other  hand, the test
we have performed is partially \textit{a posteriori} in the sense that
the short  durations of the first  few $z\gtrsim6$ bursts was  one of the
motivating factors for  conducting this study in the  first place, and
clearly it will  require a larger sample in  future to make completely
robust statements.\par

We note that 31\% (5/16) of bursts in our bright simulated sample were
easily detectable at  $z>10$. If such a population  exists at high red
shift instruments like the \textit{Swift}/BAT can and may already have
detected them although we were  unable to follow-up the detection with
a measurement of the redshift.\par

\section*{Acknowledgements}

We would  like to  thank the referee  for their useful  comments. This
work   is  supported   at   the  University   of   Leicester  by   the
STFC. P.A.E. acknowledges support from UKSA.

\begin{bibliography}{bright_sims_rev}
  \bibliographystyle{mn2e}

\begin{thebibliography}{}

\bibitem[\protect\citeauthoryear{{Barthelmy} et~al.,}{{Barthelmy}
  et~al.}{2005}]{2005SSRv..120..143B}
{Barthelmy} S.~D.,  et~al., 2005, \ssr, 120, 143

\bibitem[\protect\citeauthoryear{{Belczynski}, {Holz}, {Fryer}, {Berger},
  {Hartmann} \& {O'Shea}}{{Belczynski} et~al.}{2010}]{2010ApJ...708..117B}
{Belczynski} K.,  {Holz} D.~E.,  {Fryer} C.~L.,  {Berger} E.,  {Hartmann}
  D.~H.,    {O'Shea} B.,  2010, \apj, 708, 117

\bibitem[\protect\citeauthoryear{{Bromberg}, {Nakar}, {Piran} \&
  {Sari}}{{Bromberg} et~al.}{2013}]{2013ApJ...764..179B}
{Bromberg} O.,  {Nakar} E.,  {Piran} T.,    {Sari} R.,  2013, \apj, 764, 179

\bibitem[\protect\citeauthoryear{{Burrows} et~al.,}{{Burrows}
  et~al.}{2005}]{2005SSRv..120..165B}
{Burrows} D.~N.,  et~al., 2005, \ssr, 120, 165

\bibitem[\protect\citeauthoryear{{Chornock}, {Berger}, {Fox}, {Lunnan},
  {Drout}, {Fong}, {Laskar} \& {Roth}}{{Chornock}
  et~al.}{2013}]{2013ApJ...774...26C}
{Chornock} R.,  {Berger} E.,  {Fox} D.~B.,  {Lunnan} R.,  {Drout} M.~R.,
  {Fong} W.-f.,  {Laskar} T.,    {Roth} K.~C.,  2013, \apj, 774, 26

\bibitem[\protect\citeauthoryear{{Chornock}, {Perley}, {Cenko} \&
  {Bloom}}{{Chornock} et~al.}{2009}]{2009GCN..9243....1C}
{Chornock} R.,  {Perley} D.~A.,  {Cenko} S.~B.,    {Bloom} J.~S.,  2009, GRB
  Coordinates Network, 9243, 1

\bibitem[\protect\citeauthoryear{{Cucchiara} et~al.,}{{Cucchiara}
  et~al.}{2011a}]{2011ApJ...736....7C}
{Cucchiara} A.,  et~al., 2011a, \apj, 736, 7

\bibitem[\protect\citeauthoryear{{Cucchiara} et~al.,}{{Cucchiara}
  et~al.}{2011b}]{2011ApJ...743..154C}
{Cucchiara} A.,  et~al., 2011b, \apj, 743, 154

\bibitem[\protect\citeauthoryear{{D'Avanzo}, {D'Elia}, {di Fabrizio} \&
  {Gurtu}}{{D'Avanzo} et~al.}{2011}]{2011GCN..11997...1D}
{D'Avanzo} P.,  {D'Elia} V.,  {di Fabrizio} L.,    {Gurtu} A.,  2011, GRB
  Coordinates Network, 11997, 1

\bibitem[\protect\citeauthoryear{{de Ugarte Postigo}, {Castro-Tirado} \&
  {Gorosabel}}{{de Ugarte Postigo} et~al.}{2011}]{2011GCN..11978...1D}
{de Ugarte Postigo} A.,  {Castro-Tirado} A.~J.,    {Gorosabel} J.,  2011, GRB
  Coordinates Network, 11978, 1

\bibitem[\protect\citeauthoryear{{Fenimore}, {Palmer}, {Galassi}, {Tavenner},
  {Barthelmy}, {Gehrels}, {Parsons} \& {Tueller}}{{Fenimore}
  et~al.}{2003}]{2003AIPC..662..491F}
{Fenimore} E.~E.,  {Palmer} D.,  {Galassi} M.,  {Tavenner} T.,  {Barthelmy} S.,
   {Gehrels} N.,  {Parsons} A.,    {Tueller} J.,  2003, in {Ricker} G.~R.,
  {Vanderspek} R.~K.,  eds, Gamma-Ray Burst and Afterglow Astronomy 2001: A
  Workshop Celebrating the First Year of the HETE Mission Vol.~662 of American
  Institute of Physics Conference Series, {The Trigger Algorithm for the Burst
  Alert Telescope on Swift}.
pp 491--493

\bibitem[\protect\citeauthoryear{{Foley}, {Chen}, {Bloom} \&
  {Prochaska}}{{Foley} et~al.}{2005}]{2005GCN..3483....1F}
{Foley} R.~J.,  {Chen} H.-W.,  {Bloom} J.,    {Prochaska} J.~X.,  2005, GRB
  Coordinates Network, 3483, 1

\bibitem[\protect\citeauthoryear{{Fynbo} et~al.,}{{Fynbo}
  et~al.}{2009}]{2009ApJS..185..526F}
{Fynbo} J.~P.~U.,  et~al., 2009, \apjs, 185, 526

\bibitem[\protect\citeauthoryear{{Gehrels} et~al.,}{{Gehrels}
  et~al.}{2004}]{2004ApJ...611.1005G}
{Gehrels} N.,  et~al., 2004, \apj, 611, 1005

\bibitem[\protect\citeauthoryear{{Gehrels} et~al.,}{{Gehrels}
  et~al.}{2006}]{2006Natur.444.1044G}
{Gehrels} N.,  et~al., 2006, \nat, 444, 1044

\bibitem[\protect\citeauthoryear{{Gendre} et~al.,}{{Gendre}
  et~al.}{2010}]{2010MNRAS.405.2372G}
{Gendre} B.,  et~al., 2010, \mnras, 405, 2372

\bibitem[\protect\citeauthoryear{{Gendre} et~al.,}{{Gendre}
  et~al.}{2013}]{2013ApJ...766...30G}
{Gendre} B.,  et~al., 2013, \apj, 766, 30

\bibitem[\protect\citeauthoryear{{Genet} \& {Granot}}{{Genet} \&
  {Granot}}{2009}]{2009MNRAS.399.1328G}
{Genet} F.,  {Granot} J.,  2009, \mnras, 399, 1328

\bibitem[\protect\citeauthoryear{{Gorbovskoy} et~al.,}{{Gorbovskoy}
  et~al.}{2012}]{2012MNRAS.421.1874G}
{Gorbovskoy} E.~S.,  et~al., 2012, \mnras, 421, 1874

\bibitem[\protect\citeauthoryear{{Greiner} et~al.,}{{Greiner}
  et~al.}{2009}]{2009ApJ...693.1610G}
{Greiner} J.,  et~al., 2009, \apj, 693, 1610

\bibitem[\protect\citeauthoryear{{Gruber} et~al.,}{{Gruber}
  et~al.}{2011}]{2011A&A...528A..15G}
{Gruber} D.,  et~al., 2011, \aap, 528, A15

\bibitem[\protect\citeauthoryear{{Kann} et~al.,}{{Kann}
  et~al.}{2011}]{2011ApJ...734...96K}
{Kann} D.~A.,  et~al., 2011, \apj, 734, 96

\bibitem[\protect\citeauthoryear{{Klebesadel}, {Strong} \&
  {Olson}}{{Klebesadel} et~al.}{1973}]{1973ApJ...182L..85K}
{Klebesadel} R.~W.,  {Strong} I.~B.,    {Olson} R.~A.,  1973, \apjl, 182, L85

\bibitem[\protect\citeauthoryear{{Kocevski} \& {Petrosian}}{{Kocevski} \&
  {Petrosian}}{2013}]{2013ApJ...765..116K}
{Kocevski} D.,  {Petrosian} V.,  2013, \apj, 765, 116

\bibitem[\protect\citeauthoryear{{Kouveliotou}, {Meegan}, {Fishman}, {Bhat},
  {Briggs}, {Koshut}, {Paciesas} \& {Pendleton}}{{Kouveliotou}
  et~al.}{1993}]{1993ApJ...413L.101K}
{Kouveliotou} C.,  {Meegan} C.~A.,  {Fishman} G.~J.,  {Bhat} N.~P.,  {Briggs}
  M.~S.,  {Koshut} T.~M.,  {Paciesas} W.~S.,    {Pendleton} G.~N.,  1993,
  \apjl, 413, L101

\bibitem[\protect\citeauthoryear{{Kr{\"u}hler} et~al.,}{{Kr{\"u}hler}
  et~al.}{2012}]{2012A&A...546A...8K}
{Kr{\"u}hler} T.,  et~al., 2012, \aap, 546, A8

\bibitem[\protect\citeauthoryear{{Levan} et~al.,}{{Levan}
  et~al.}{2013}]{2013arXiv1302.2352L}
{Levan} A.~J.,  et~al., 2013, ArXiv e-prints

\bibitem[\protect\citeauthoryear{{Levan}, {Perley}, {Tanvir} \&
  {Cucchiara}}{{Levan} et~al.}{2012}]{2012GCN..13802...1L}
{Levan} A.~J.,  {Perley} D.~A.,  {Tanvir} N.~R.,    {Cucchiara} A.,  2012, GRB
  Coordinates Network, 13802, 1

\bibitem[\protect\citeauthoryear{{L{\"u}}, {Zhang}, {Liang}, {Zhang} \&
  {Sakamoto}}{{L{\"u}} et~al.}{2012}]{2012arXiv1211.1117L}
{L{\"u}} H.-J.,  {Zhang} B.,  {Liang} E.-W.,  {Zhang} B.-B.,    {Sakamoto} T.,
  2012, ArXiv e-prints

\bibitem[\protect\citeauthoryear{{McBreen} et~al.,}{{McBreen}
  et~al.}{2010}]{2010A&A...516A..71M}
{McBreen} S.,  et~al., 2010, \aap, 516, A71

\bibitem[\protect\citeauthoryear{{Meegan}, {Fishman}, {Wilson}, {Horack},
  {Brock}, {Paciesas}, {Pendleton} \& {Kouveliotou}}{{Meegan}
  et~al.}{1992}]{1992Natur.355..143M}
{Meegan} C.~A.,  {Fishman} G.~J.,  {Wilson} R.~B.,  {Horack} J.~M.,  {Brock}
  M.~N.,  {Paciesas} W.~S.,  {Pendleton} G.~N.,    {Kouveliotou} C.,  1992,
  \nat, 355, 143

\bibitem[\protect\citeauthoryear{{Nakar}}{{Nakar}}{2007}]{2007PhR...442..166N}
{Nakar} E.,  2007, \physrep, 442, 166

\bibitem[\protect\citeauthoryear{{Norris} \& {Bonnell}}{{Norris} \&
  {Bonnell}}{2006}]{2006ApJ...643..266N}
{Norris} J.~P.,  {Bonnell} J.~T.,  2006, \apj, 643, 266

\bibitem[\protect\citeauthoryear{{Norris}, {Marani} \& {Bonnell}}{{Norris}
  et~al.}{2000}]{2000ApJ...534..248N}
{Norris} J.~P.,  {Marani} G.~F.,    {Bonnell} J.~T.,  2000, \apj, 534, 248

\bibitem[\protect\citeauthoryear{{O'Meara}, {Chen} \& {Prochaska}}{{O'Meara}
  et~al.}{2010}]{2010GCN..11089....1O}
{O'Meara} J.,  {Chen} H.~W.,    {Prochaska} J.~X.,  2010, GRB Coordinates
  Network, 11089, 1

\bibitem[\protect\citeauthoryear{{Paciesas} et~al.,}{{Paciesas}
  et~al.}{1999}]{1999ApJS..122..465P}
{Paciesas} W.~S.,  et~al., 1999, \apjs, 122, 465

\bibitem[\protect\citeauthoryear{{Paciesas} et~al.,}{{Paciesas}
  et~al.}{2012}]{2012ApJS..199...18P}
{Paciesas} W.~S.,  et~al., 2012, \apjs, 199, 18

\bibitem[\protect\citeauthoryear{{Page} et~al.,}{{Page}
  et~al.}{2007}]{2007ApJ...663.1125P}
{Page} K.~L.,  et~al., 2007, \apj, 663, 1125

\bibitem[\protect\citeauthoryear{{Racusin} et~al.,}{{Racusin}
  et~al.}{2008}]{2008Natur.455..183R}
{Racusin} J.~L.,  et~al., 2008, \nat, 455, 183

\bibitem[\protect\citeauthoryear{{Reichart}, {Lamb}, {Fenimore},
  {Ramirez-Ruiz}, {Cline} \& {Hurley}}{{Reichart}
  et~al.}{2001}]{2001ApJ...552...57R}
{Reichart} D.~E.,  {Lamb} D.~Q.,  {Fenimore} E.~E.,  {Ramirez-Ruiz} E.,
  {Cline} T.~L.,    {Hurley} K.,  2001, \apj, 552, 57

\bibitem[\protect\citeauthoryear{{Sakamoto} et~al.,}{{Sakamoto}
  et~al.}{2008}]{2008ApJS..175..179S}
{Sakamoto} T.,  et~al., 2008, \apjs, 175, 179

\bibitem[\protect\citeauthoryear{{Sakamoto} et~al.,}{{Sakamoto}
  et~al.}{2011}]{2011ApJS..195....2S}
{Sakamoto} T.,  et~al., 2011, \apjs, 195, 2

\bibitem[\protect\citeauthoryear{{Scargle}}{{Scargle}}{1998}]{1998ApJ...504..4%
05S}
{Scargle} J.~D.,  1998, \apj, 504, 405

\bibitem[\protect\citeauthoryear{{Tanvir} et~al.,}{{Tanvir}
  et~al.}{2009}]{2009Natur.461.1254T}
{Tanvir} N.~R.,  et~al., 2009, \nat, 461, 1254

\bibitem[\protect\citeauthoryear{{Tanvir} et~al.,}{{Tanvir}
  et~al.}{2012}]{2012GCN..13348...1T}
{Tanvir} N.~R.,  et~al., 2012, GRB Coordinates Network, 13348, 1

\bibitem[\protect\citeauthoryear{{Vreeswijk} et~al.,}{{Vreeswijk}
  et~al.}{2007}]{2007A&A...468...83V}
{Vreeswijk} P.~M.,  et~al., 2007, \aap, 468, 83

\bibitem[\protect\citeauthoryear{{Wiersema}, {Tanvir}, {Vreeswijk}, {Fynbo},
  {Starling}, {Rol} \& {Jakobsson}}{{Wiersema}
  et~al.}{2008}]{2008GCN..7517....1W}
{Wiersema} K.,  {Tanvir} N.,  {Vreeswijk} P.,  {Fynbo} J.,  {Starling} R.,
  {Rol} E.,    {Jakobsson} P.,  2008, GRB Coordinates Network, 7517, 1

\bibitem[\protect\citeauthoryear{{Willingale} et~al.,}{{Willingale}
  et~al.}{2007}]{2007ApJ...662.1093W}
{Willingale} R.,  et~al., 2007, \apj, 662, 1093

\bibitem[\protect\citeauthoryear{{Willingale}, {Genet}, {Granot} \&
  {O'Brien}}{{Willingale} et~al.}{2010}]{2010MNRAS.403.1296W}
{Willingale} R.,  {Genet} F.,  {Granot} J.,    {O'Brien} P.~T.,  2010, \mnras,
  403, 1296

\bibitem[\protect\citeauthoryear{{Xu}, {Fynbo}, {Tanvir}, {Hjorth}, {Leloudas},
  {Malesani}, {Jakobsson}, {Wilson} \& {Andersen}}{{Xu}
  et~al.}{2009}]{2009GCN..10053...1X}
{Xu} D.,  {Fynbo} J.~P.~U.,  {Tanvir} N.~R.,  {Hjorth} J.,  {Leloudas} G.,
  {Malesani} D.,  {Jakobsson} P.,  {Wilson} P.~A.,    {Andersen} J.,  2009, GRB
  Coordinates Network, 10053, 1

\bibitem[\protect\citeauthoryear{{Zhang} et~al.,}{{Zhang}
  et~al.}{2009}]{2009ApJ...703.1696Z}
{Zhang} B.,  et~al., 2009, \apj, 703, 1696

\bibitem[\protect\citeauthoryear{{Zhang}, {Zhang}, {Liang}, {Gehrels},
  {Burrows} \& {M{\'e}sz{\'a}ros}}{{Zhang} et~al.}{2007}]{2007ApJ...655L..25Z}
{Zhang} B.,  {Zhang} B.-B.,  {Liang} E.-W.,  {Gehrels} N.,  {Burrows} D.~N.,
  {M{\'e}sz{\'a}ros} P.,  2007, \apjl, 655, L25

\end{thebibliography}
\end{bibliography}

\appendix
\section{Method 1 detailed description}

To simulate  BAT light curves,  we obtained the event  lists extracted
for  each burst  using {\sc  batgrbproduct}.  The  additional required
information  included the BAT  Ancillary Response  File (ARF)  and the
actual observed redshift of the GRB in question, $z_{\rm orig}$.\par

As well as the number of  source counts within each bin being reduced,
the    duration   over    which    they   are    received   is    time
dilated.  Specifically,  each simulated  bin  now  has  a duration  of
64$\zfac$\,ms. As the  arrival of each photon is  a Poisson process we
could  not simply  derive which  fraction of  the new  bin  size falls
within   a  single  64\,ms   bin.  Instead   we  took   the  extracted
15$\zfac$--25$\zfac$ light  curve and  calculated the total  number of
counts observed  in each bin (by  correcting for the bin  size and the
number   of  fully   illuminated  detectors).   We  also   required  a
background. To  find this we  looked at the  RMS scatter on  the light
curve,  and  added  an offset  equal  to  the  square of  this  value,
appropriate for Poisson noise.\par

Having  the  total  number of  counts,  we  then  used $Q$  (where  by
definition, $Q\leqslant1$)  to consider  whether each would  remain in
the light curve when the source was moved to $z_{\rm sim}$.  To do so,
we  generated a  random  number $R_{1}$  from  a uniform  distribution
ranging between 0 and 1.  The  value of $Q$ also took this range, with
1 corresponding to  simulating the light curve at  the redshift it was
observed at.  Any count for which $R_{1}$ $\leqslant$ $Q$ was retained
in the light curve.\par

At this point we  re--binned the light curve to the original 64\,ms. 
To  do so, we generated  another random number,  $R_{2}$. This was
again drawn from a uniform distribution, ranging between 0 and 1. This
number corresponded to the fraction  of bin duration which had elapsed
when the count arrived. The time of each event, $t_{\rm evt}$, is given as
expressed in Equation \ref{eq:evttime},  where $t_{\rm st}$ is the time at
which the bin begins.\par

\begin{equation}
  t_{\rm evt} = \zfac\left( t_{\rm st} + 0.064R_{2} \right).
  \label{eq:evttime}
\end{equation}

Given a time for each  count, they were re--sampled to 64\,ms temporal
binning, the  background offset was  removed, and the light  curve was
returned  to the  units  of cts.s$^{{\rm  -1}}$.det$^{-1}$.  This  new
light  curve now  contained the  correct number  of source  counts. By
scaling the total number of counts  by a factor of $Q$, we also scaled
the background was  by the same quantity.  This  meant the total range
in the fluctuations  in the new background subtracted  light curve was
underestimated, and therefore an additional noise component was added.
To correct  the background, we  calculated the variance  on quiescent,
non--slew times of  both the transformed light curve  and the original
observed 15--25\,keV light curve.   The latter was also extracted from
the available BAT data using  the {\sc batbinevt} routine. We compared
these two  variances, and found  the difference, as shown  in Equation
\ref{eq:vars}.\par

\begin{equation}
  \sigma_{\rm diff}^{2} = \sigma_{\rm orig}^{2} - \sigma_{\rm sim}^{2}.
  \label{eq:vars}
\end{equation}

A further  series of  random numbers was  then generated.   These were
taken  from a  Gaussian distribution  with mean  of zero  and standard
deviation $\sigma_{\rm diff}$.   We drew one random number  per bin in
the 64  ms simulated light curve.   Each of these  random numbers were
added to their associated bin to increase the scatter on the simulated
light curve  to the level as  seen in the observed  15--25 \,keV light
curve.   Having added  this  additional scatter,  the resultant  light
curve contained  the correct count rate and  noise characteristics due
to both the source and background.\par

\section{Simulating noise for Method 2}

In   this  case  the   simulated  light   curves  initially   have  no
noise\footnote{strictly  speaking  the  noise  in the  original  light
  curves  does affect the  model fits,  but since  our sample  are all
  detected  bursts, and  the  fits effectively  smooth  the data,  the
  residual  noise  effect  is   minor  (particularly  for  the  bright
  subset).}, so this must be added  in a realistic way.  To do so, the
light curve was  converted into a photon count rate  per bin, then for
each  bin  a Poisson  distribution  was  randomly  sampled, using  the
modelled   number  of  counts   as  the   expectation  value   of  the
distribution. This  random number was  then taken to be  the simulated
number of  counts. This accounts  for noise due  to the source  and to
this  we must  add  a  background contribution.   To  achieve this  we
considered the observed light curve data  from BAT in each of the four
bands at the time during which  the GRB was defined as active. We then
considered  the RMS  value of  errors  on the  light curve  (excluding
bright  active regions)  to  find  an average  value  of noise.   This
average  noise was  used  as  the standard  deviation  for a  Gaussian
distribution which had a mean of zero.  Random numbers were drawn from
this distribution and added to each  bin of the four BAT channels.  In
principle  the  background variations  could  be  both shot-noise  and
variations in,  for example, astrophysical  sources in the  field. Our
procedure accounts  for any such variations that  are reasonably fast,
but not any  slower variations (which would be  correlated from bin to
bin).\par

\label{lastpage}
\end{document}